\documentclass[sigplan,10pt]{acmart}
\makeatletter
\def\@ACM@checkaffil{}

\acmBadgeR[https://www.acm.org/publications/policies/artifact-review-and-badging-current]{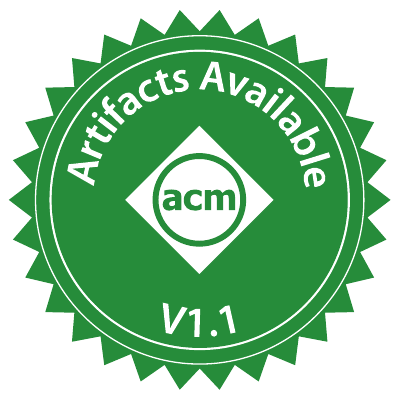}
\acmBadgeR[https://www.acm.org/publications/policies/artifact-review-and-badging-current]{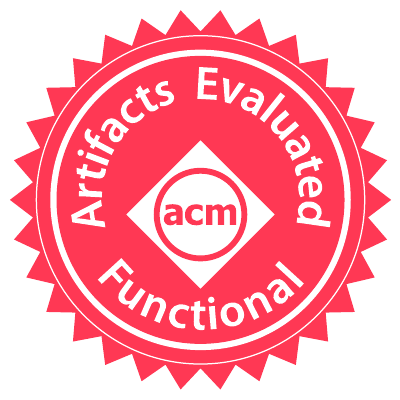}
\acmBadgeR[https://www.acm.org/publications/policies/artifact-review-and-badging-current]{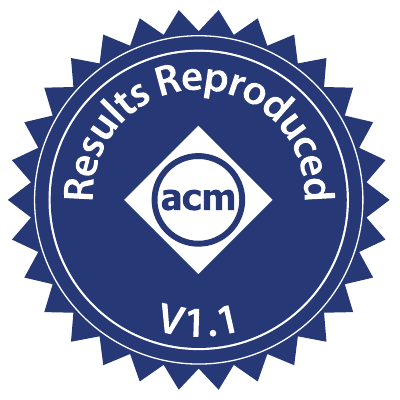}
\usepackage{balance}


\copyrightyear{2027}
\acmYear{2027}
\setcopyright{cc}
\setcctype{by}
\acmConference[EuroSys '27]{22nd European Conference on Computer Systems}{April 19--23, 2027}{Rabat, Morocco}
\acmBooktitle{22nd European Conference on Computer Systems (EuroSys '27), April 19--23, 2027, Rabat, Morocco}
\acmDOI{10.1145/3842654.3848552}
\acmISBN{979-8-4007-2971-3/2027/04}
\makeatother
\usepackage{multirow}
\usepackage{fvextra}
\usepackage{makecell}
\usepackage{etoolbox}
\usepackage{algorithm}
\usepackage{algpseudocode}
\usepackage{graphicx}
\usepackage[normalem]{ulem}
\usepackage{tabularx}
\usepackage{booktabs}
\usepackage{adjustbox}
\usepackage{array}
\usepackage{enumitem}
\usepackage{tcolorbox,tikz}
\usepackage{pifont}
\usepackage{multicol}
\usepackage{physics}
\usepackage{upgreek}
\usepackage{xurl}

\definecolor{aliceblue}{rgb}{0.94, 0.97, 1.0}
\tcbset{
  width=1\columnwidth,
  boxrule=1pt,
  colback=aliceblue,
  arc=1pt,
  left=2pt,
  right=2pt,
  boxsep=0.5pt
}

\definecolor{F0FAFF}{HTML}{F0FAFF}
\definecolor{wcBg}{HTML}{FFF1F1}

\newcommand*\circled[1]{%
  \tikz[baseline=(char.base)]{
    \node[shape=circle, fill=black, draw, inner sep=0.1pt] (char) {%
      \color{white}
      {\fontsize{8pt}{9pt}\selectfont #1}%
    };
  }
}

\AtBeginDocument{%
  \hypersetup{
    colorlinks=true,
    citecolor=blue,
    linkcolor=blue,
    urlcolor=blue
  }%
}

\newcommand{\ignore}[1]{}

\definecolor{codegray}{gray}{0.97}
\newcommand{\design}{\textsc{IRIS}}
\newcommand{\Input}{\State\textbf{Input:} }
\newcommand{\Output}{\State\textbf{Output:} }
\newcommand{\ums}{UMS}
\newcommand{\ees}{EES}
\AtBeginDocument{%
  \hypersetup{
    colorlinks=true,
    citecolor=blue,
    linkcolor=blue,
    urlcolor=blue
  }%
}
\begin{document}
\title{\design{}: A Compiler for Distributed Quantum Systems}
\author{Won Joon Yun}
\email{wonjoon.yun@utexas.edu}
\orcid{0000-0003-0405-8843}
\affiliation{%
  \institution{The University of Texas at Austin}
}
\author{Dhilan Nag}
\email{dhilannag@utexas.edu}
\orcid{0009-0003-6147-3012}
\affiliation{%
  \institution{The University of Texas at Austin}
}

\author{Sneha Ballabh}
\email{snehaballabh@utexas.edu}
\orcid{0000-0002-2897-4208}
\affiliation{%
  \institution{The University of Texas at Austin}
}

\author{Jiapeng Zhao}
\email{penzhao2@cisco.com}
\orcid{0000-0002-7851-4648}
\affiliation{%
  \institution{Cisco Quantum Lab}
}

\author{Eneet Kaur}
\email{ekaur@cisco.com}
\orcid{0000-0003-3407-6284}
\affiliation{%
  \institution{Cisco Quantum Lab}
}

\author{Poulami Das}
\email{poulami.das@utexas.edu}
\orcid{0000-0002-5811-6108}
\affiliation{%
  \institution{The University of Texas at Austin}
}
\begin{abstract}
Distributed Quantum Computers (DQCs) enable large system sizes by connecting smaller chips via photonic interconnects. DQCs use teleportation to relocate qubits and execute CNOTs between qubits on different chips. 
But, non-local CNOTs are 4.3-7.7$\times$ slower and 4$\times$ more error-prone than local CNOTs within a chip, which degrades program fidelities. 
Compilers group CNOTs with overlapping qubits into \textit{blocks} and collectively optimize teleportations for each block. However, block-level scheduling suffers from two key drawbacks. \textit{First}, they lack lookahead ability between blocks because they select the best schedule for a block before proceeding. So, they cannot assess the impact of a teleportation on future blocks. Our studies show that naively expanding the lookahead window to include subsequent blocks does not address the issue.   
\textit{Second}, they do not schedule future block operations or teleportations they need unless preceding blocks are fully scheduled, introducing delay and latency overheads.

We propose \textit{\design{}}, a DQC compiler that addresses these limitations using two key insights: \textit{Utility-driven Lookahead With Multi-Candidate Block Scheduling (\ums)} and \textit{EPR-Capacity-Aware Early Scheduling (\ees{})}. \ums{} schedules a block by considering only useful future blocks in its lookahead window. A future block has utility if it shares overlapping qubits with the current block being scheduled. \ums{} also maintains multiple schedules during compilation, allowing it to defer commitment to globally sub-optimal schedules early. \ees{} enables \design{} to schedule future operations and their relocations early if EPR resources are available for teleportations. 
Our evaluations show that \design{} reduces teleportations by 34\% on average and by up to 65\%; and latency by 2$\times$ on average and by up to 2.9$\times$ compared to the state-of-the-art DQC compiler.
\end{abstract}

\hypersetup{urlcolor=black}
\renewcommand{\shortauthors}{}
\maketitle
\hypersetup{urlcolor=blue}

\section{Introduction}

Distributed Quantum Computers (DQCs) overcome low yield and fidelity concerns of large monolithic chips by connecting smaller chips via photonic interconnects~\cite{DQC-ARQUIN,DATACENTER-ARCH,DQC-Arch,vanmeter2006distributed,DQC-Oh2023Distributed,DQC-Chandra2024Network}. DQCs offer a pathway to scale system sizes and recent demonstrations~\cite{Li2024Highrate,DQC-Distributed2025Main,DQC-Experiment1,CarreraVazquez2024,Wei2025UDBQC} have led to their inclusion in industrial and academic road-maps~\cite{ionq_roadmap_2024,pasqal_roadmap_2025,ibmqroadmap,aghaee2025scaling}. Two-qubit CNOT or \textit{entanglement} operations, critical to attain quantum advantage, between qubits on a chip are identical to monolithic systems. In contrast, non-local CNOTs between different chips use \textit{teleportations} to relocate required qubits to the same chip, after which the CNOT is executed locally~\cite{TeleData,TeleGate}. Unfortunately, non-local CNOTs exhibit 4$\times$ higher error-rates and take 4.3-7.7$\times$ longer than local CNOTs~\cite{Li2024Highrate,Bluvstein2024}. This degrades program fidelity by increasing the vulnerability to errors and decoherence.

DQC compilers improve program fidelity by selecting qubit relocation paths that minimize teleportations and maximize concurrency~\cite{cuomo2023optimized,xu2025optimizing}. The GP compiler selects the shortest qubit relocation paths for each non-local CNOT~\cite{ferrari2020}. AutoComm~\cite{DQC-Wu2022AutoComm} selects paths collectively for a \textit{block} of non-local {CNOTs} between two chips, if they have a qubit in common, such that the total number of relocations needed to execute the block is minimized. QuComm~\cite{DQC-Wu2023QuComm} extends AutoComm to form even larger blocks by considering non-local CNOTs spread over multiple chips, as long as sufficient EPR capacity is available on a chip to execute them. The \textit{EPR capacity} denotes the number of external qubits a chip can hold at any given time and is mandatory for teleportations. Existing compilers optimize the schedule for each block by selecting relocation paths that maximize the number of local CNOT gates within the block and minimize the number of teleportations required for the non-local gates collectively.

\begin{figure*}[t!]
\centering\includegraphics[width=\textwidth]{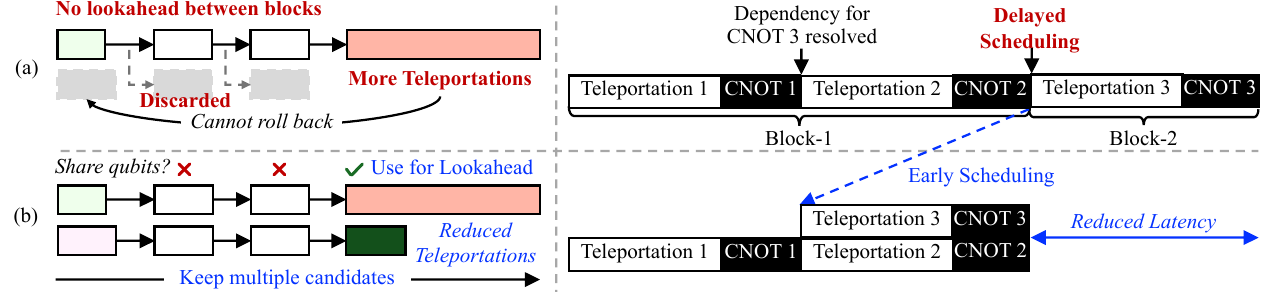}
\caption{
(a) Current compilers have limited lookahead capability between blocks and commit to the schedule with fewest teleportations for each block while discarding the rest before proceeding to the next block.
They also defer scheduling future gates until preceding blocks complete.
(b) \design{} considers useful future blocks in its lookahead window and retains multiple candidate schedules. It also executes future operations and their teleportation early if sufficient EPR capacity is available.
}
\label{fig:fig-1}
\end{figure*}

Block-level scheduling, however, has two key drawbacks. \textit{First}, it lacks lookahead ability between blocks. As teleportations relocate qubits, any teleportation impacts the scheduling of all future gates and teleportations both. Current compilers only minimize teleportations for each block but do not assess how these relocations impact future blocks. So, for a program with $N$ blocks, $B_1$ to $B_N$, they cannot evaluate how qubit displacements in $B_1$ impact teleportations in $B_2$ to $B_N$. \textit{Ideally}, a compiler must use all future blocks in the \textit{lookahead window} and assess this impact, but it is not feasible due to exponential complexity. In practice, it may only be able to consider $F$ future blocks ($F{\ll}N$).
Also, unlike monolithic compilers where only considering the next $F$ gates suffices for lookahead heuristics, our studies show that simply using the next $F$ blocks does not work for DQCs. To show this, we consider 
a program with 20 random blocks. The small size allows us to derive the optimal schedule. By using a lookahead window with the next 10 blocks while scheduling each block, QuComm reduces teleportations from 21 to 16. However, it is still higher than the optimal (13). This is because a chip typically holds many qubits which means several blocks comprise only local CNOTs; and consecutive teleportations on a qubit are often several blocks apart (distance$>$F). 
Our studies using a QAOA program with 57K CNOTs grouped into 17K blocks show that there are $\sim$260 CNOTs on average between consecutive relocations on a qubit. With an average block size of $\sim$3 CNOTs here, these two relocations are $\sim$89 blocks apart. Thus, a lookahead window with only a few future blocks does not suffice. Committing to a block schedule early also limits lookahead performance because a schedule that is optimal for a block may increase teleportations in future, as shown in Figure~\ref{fig:fig-1}(a). Alternate schedules that may reduce teleportations in future are discarded by the compiler. 

\begin{figure*}[t]
\centering\includegraphics[width=\textwidth]{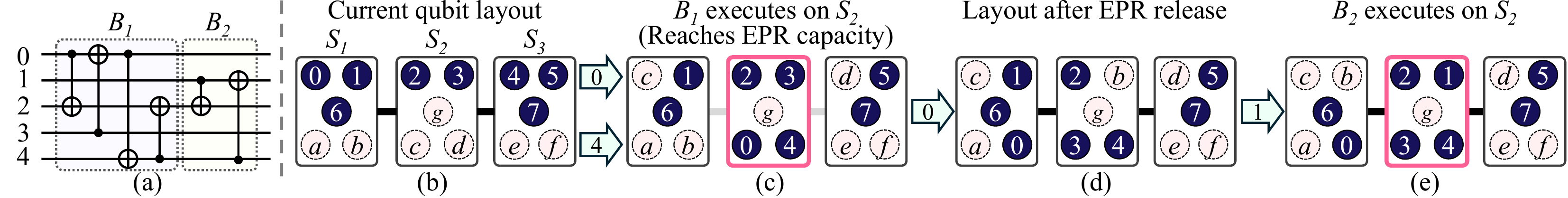}
    \caption{
    Compilation of (a) a circuit on (b) a 3-chip DQC using state-of-the-art DQC compiler QuComm~\cite{DQC-Wu2023QuComm}. QuComm groups the CNOTs into two blocks, $B_1$ and $B_2$. (c) It selects $S_2$ and relocates qubits 0 and 4 in parallel to execute $B_1$ on $S_2$, exhausting its EPR resources. (d) To free EPRs, QuComm relocates qubit 0 from $S_2$ to $S_1$. (e) It then relocates qubit 1 to $S_2$ for $B_2$.}
    \label{fig:fig-2}
\end{figure*}

The second limitation of block-level scheduling is that a block is scheduled only after \textit{all} preceding blocks have been scheduled. Also, teleportations are only scheduled when the compiler encounters a non-local gate (\textit{on-demand scheduling}). This causes future gates and any teleportations they require to wait to be scheduled even if data dependencies have been resolved earlier, increasing latency.
For example, if block $B_1$ unlocks a data dependency for a gate in block $B_2$, the gate and its teleportations are not scheduled until the compiler has scheduled all operations up to that gate in $B_2$, as shown in Figure~\ref{fig:fig-1}(a). 
Our studies show that 45.4\% of teleportations wait for 7.5\,ms on average in the same QAOA program above.

We propose {\em \design{}}, a compiler that addresses these issues by using two key insights. \textit{First}, we design \textit{\underbar{U}tility-Driven Lookahead With \underbar{M}ulti-Candidate Block \underbar{S}cheduling (UMS)}. 
To optimize the lookahead heuristic, \ums{} leverages a utility-driven approach instead of naively considering $N$ subsequent blocks. 
A future block has \textit{utility} in the lookahead heuristic of the current block being scheduled, only if both have overlapping qubits.
This enables \design{} to optimize teleportation routes with a small lookahead window without drastically increasing compilation complexity. However, evaluating the impact of a current block schedule on the teleportation overheads in future blocks is non-trivial because the overhead of an $i^{\text{th}}$ future block can only be determined by actually scheduling it starting with qubit positions at the end of the $(i-1)^{\text{th}}$ block. Moreover, once a schedule for the current block is already committed to the final executable, the compiler cannot evaluate alternatives that may have led to fewer teleportations in the future. To address this issue, \ums{} retains multiple schedules for the current block. Then, starting with each schedule and corresponding qubit positions, \design{} schedules subsequent blocks, eventually building a tree of solutions. For each non-local CNOT, \ums{} selects teleportation routes with minimal relocations.
To keep the complexity tractable, \design{} prunes the tree and retains only Top-$w$ solutions, where $w$ is the maximum number of solutions considered.   

\textit{Second}, we propose \textit{\underline{E}PR-capacity-aware \underline{E}arly \underline{S}cheduling (\ees{})} that schedules future CNOTs and any teleportations they require, if EPR capacity is available, as early as possible. Early scheduling is non-trivial because dependent operations are often many blocks apart and form long chains. Scheduling teleportations early alters EPR resource distributions and may increase teleportation costs of intermediate operations.
For example, if a gate in block $B_1$ resolves the dependency for a CNOT in $B_{5}$, scheduling the $B_{5}$ gate early may change EPR capacities, increasing relocation costs for operations pending in $B_1$ to $B_4$. To address this, \ees{} schedules a program, collects the set of gates $E$ that can be scheduled early by tracing instruction dependency chains, and tracks EPR capacities and qubit positions post each operation. Next, \ees{} schedules each operation in $E$ earlier than its current timeline if it does not increase the number of teleportations required. \ees{} ensures that such early scheduling teleportations does not fully consume the EPR capacity of a chip; else, some EPR resources would need to be freed for downstream operations.
Figure~\ref{fig:fig-1}(b) illustrates how \design{} yields fewer teleportations overall by employing utility-driven lookahead and retaining multiple candidate schedules. It also reduces latency by enabling EPR-capacity-aware early scheduling.

Our evaluations using representative benchmarks and DQCs show that \design{} reduces teleportations and latencies by $34\%$ and $2\times$ on average. \design{} is available at: {\color{blue}{\href{https://doi.org/10.5281/zenodo.22152933}{IRIS-Code}}}.

\vspace{0.05in}
To that end, this paper makes the following contributions:
\begin{enumerate}[leftmargin=2pt, itemindent=2pt, labelsep=3pt,itemsep=1pt]
\item We design \textit{\design{}}, a compiler to reduce the overheads of teleportations in Distributed Quantum Computers (DQCs).

\item We propose \textit{\underbar{U}tility-Driven Lookahead with \underbar{M}ulti-Candidate Block \underbar{S}cheduling (\ums{})}, which identifies the relevant future blocks that have utility in the lookahead heuristics of each block and manages multiple schedules at any given time to prevent convergence on sub-optimal schedules. 

\item We propose \textit{\underbar{E}PR-capacity-aware \underbar{E}arly \underbar{S}cheduling (\ees{})} that schedules future operations and any relocations required by them as soon as their data dependencies are resolved.

\end{enumerate}

\section{Background}\subsection{Distributed Quantum Computers (DQCs)}
DQCs connect smaller quantum chips via photonic channels. Typically, chips have nearest-neighbor connectivity because all-to-all connectivity is expensive and reduces fidelity~\cite{DQC-Scaling2022Smith,zhang+:MECH,DQC-Oh2023Distributed,DQC-Chandra2024Network,Li2024Highrate,DQC-Error2024lukin-group}. 
Non-local CNOTs between qubits on different chips use teleportations~\cite{TeleData,TeleGate}. 
Teleportations require EPR pairs, for which qubits must be reserved. 
We provide more details on EPR pair generation in Appendix~\ref{appendix:epr_generation}.
Figure~\ref{fig:fig-2}(b) shows a DQC with five qubits per chip, blue qubits are \textit{compute} qubits ($0$--$7$), while the others ($a$--$g$) are communication qubits used to hold external qubit's quantum state or remotely control a qubit. We consider two types of teleportations:

\vspace{0.05in}
\noindent \textbf{RELOCATE}~\cite{TeleData}: or \textit{state teleportation} transfers the state of a qubit from one chip to another. 
For a non-local CNOT between adjacent chips, a RELOCATE moves one of the qubits to the other chip, after which the CNOT becomes local. For non-adjacent chips, multiple RELOCATEs are used to bring both qubits on the same chip. RELOCATEs displace qubits and the number of external qubits a chip can hold equals the number of communication qubits or \textit{EPR capacity}.

\vspace{0.05in}
\noindent \textbf{Remote CNOT (Re-CNOT)}~\cite{TeleGate}: or  \textit{gate teleportation}
enables non-local CNOTs between adjacent chips without relocating a qubit. 
For example, to perform a CNOT between qubits $0$ and $2$ in Figure~\ref{fig:fig-2}(b), we may RELOCATE $0$ to $S_2$ using EPR pair $b$-$c$. Qubit $c$ holds the state of $0$ after the RELOCATE. If we were to use a Re-CNOT instead, qubit $0$ remains on $S_1$. The EPR pair is still required for the Re-CNOT as one of its qubits serves as the proxy control for the CNOT. 
For a CNOT between qubits $0$ and $4$, we must either relocate both qubits to $S_2$ or relocate $0$ to $S_3$ via $S_2$ because $S_1$ and $S_3$ are not connected, thus requiring multiple RELOCATEs.

We provide details of these operations in Appendix~\ref{appendix:teleportations}.

\vspace{-0.1in}
\subsection{The Problem: Teleportation Overheads}
RELOCATEs and Re-CNOTs take 4.3$\times$ to 7.7$\times$ longer and have 4$\times$ higher error rates than local CNOTs~\cite{Bluvstein2024,Li2024Highrate}. 
Although, programs often use far fewer teleportations than local CNOTs, the former dominates latency. 
Our studies with a 120-qubit QAOA program on a 2$\times$2 DQC show that even though only 6\% of operations are teleportations, they consume 50\% of the latency because teleportations have much lower concurrency (1.1) than local CNOTs (3.9) because teleportations are often serialized due to the limited connectivity between chips on DQCs and availability of EPR resources.

\vspace{-0.1in}
\subsection{Prior Works on DQC Compilation}\label{subsec:prior_works}
Compilers minimize teleportations and maximize concurrency. 
Teleportations displace qubits, and each displacement can make subsequent CNOTs local or non-local.
The GP compiler schedules each non-local CNOT independently along the shortest relocation path~\cite{ferrari2020}. 
This minimizes teleportation for the CNOT but ignores how the displacement affects subsequent CNOTs.
AutoComm addresses this by scheduling multiple CNOTs together, 
so that one displacement can benefit them all. AutoComm groups non-local CNOTs that share a qubit and span two chips into a \textit{block}~\cite{DQC-Wu2022AutoComm}, and schedules the block as a whole. 
For example, CNOT 0,2 and CNOT 3,0 in Figure~\ref{fig:fig-2}(b) share qubit 0. Scheduling them individually could cost up to two RELOCATEs, while scheduling 
them as a block costs one, because relocating qubit 0 to $S_2$ 
makes both CNOTs local.
QuComm~\cite{DQC-Wu2023QuComm} extends AutoComm further. It forms larger blocks spanning multiple chips, as long as the block does not exceed the EPR capacity of a chip. 
For each block, QuComm constructs a  \textit{block schedule}, a sequence of teleportations and local CNOTs that executes all gates in the block. QuComm generates this block schedule in two steps:

\vspace{0.05in}
\noindent \textbf{Block Formation:} QuComm groups non-CNOTs into blocks, such that the EPR capacity required to execute each block is available on a single chip. It also merges local CNOTs into the blocks so that RELOCATEs do not make them non-local.

\vspace{0.05in}
\noindent \textbf{Block Scheduling:} For each non-local CNOT, QuComm 
decides whether to use RELOCATE or Re-CNOT. If RELOCATEs are used, it identifies the qubits to relocate and the chip to relocate to, such that the total number of RELOCATEs required for the current gate and the remaining gates in the block are minimized. If a target chip has only one communication qubit (such as communication qubit $g$ on chip $S_2$ in Figure~\ref{fig:fig-2}(c)), QuComm evicts some of its current qubits to other chips using an operation called \textit{EPR release}. For this, it prioritizes qubits whose relocations assist future blocks.

We explain QuComm using Figure~\ref{fig:fig-2}. QuComm groups the CNOTs into blocks $B_1$ and $B_2$. For $B_1$, QuComm selects chip $S_2$ and relocates qubits $0$ and $4$ in parallel, requiring fewest RELOCATEs and yielding maximum concurrency. Scheduling on $S_1$ costs four RELOCATEs ($2$, $3$, and $4$), two of which are serialized (relocating $4$ to $S_1$ via $S_2$), whereas scheduling on $S_3$ requires three RELOCATEs ($0$, $2$), two of which are serialized (relocating $0$ to $S_3$ via $S_2$). 
Block $B_2$ cannot be scheduled because EPR resources on $S_2$ are exhausted. QuComm relocates qubit $0$ to $S_1$ to release EPR capacity, after which qubit $1$ relocates to $S_2$, where the block executes.
\section{Limitations of Prior Works and Insights}
Block-level scheduling in prior works suffers from two issues.
\subsection{No Lookahead Between Blocks}
A teleportation impacts relocation costs of all future teleportations, whether they belong to non-local gates within the same block that is being scheduled or future blocks. 
For example, if qubit $q$ is relocated in block $B_0$, it impacts the scheduling of all remaining gates in $B_0$. Moreover, if the same qubit must be relocated again while scheduling block $B_{10}$, then the relocation costs of $B_{10}$ depend on the position of $q$ based on teleportations used in $B_0$ scheduling. Thus, while selecting teleportations for $q$ during $B_0$ scheduling, we must satisfy two objectives: (1)~minimize relocation costs of the current block $B_0$ (\textit{\textbf{Objective-1}}) and (2)~minimize relocation costs of future blocks (such as $B_{10}$) (\textit{\textbf{Objective-2}}). 
An \textbf{\textit{ideal lookahead}} DQC compiler must optimize for both objectives for all teleportations. But, this is 
impractical as its complexity scales exponentially with the size of programs.
Existing compilers, like QuComm, employ \textit{low-cost lookahead heuristics} that only optimize for Objective-1 to schedule a block. They commit this schedule to the executable and only then proceed to the next block. Thus, they lack \textit{lookahead ability between blocks} and only optimize for Objective-2.

Lookahead heuristics, well studied for monolithic systems, use a set of \textit{subsequent} gates in the lookahead window while scheduling a gate~\cite{li2018tackling}. 
But naively adopting similar approaches, where the lookahead window comprises \textit{the next $F$ future blocks} is insufficient in DQCs for two reasons. 
\textit{First}, programs comprise several blocks with only local gates in between two consecutive teleportations on the same qubit. 
Including these \textit{local-only} blocks in the lookahead window offer no practical benefits.
\textit{Second}, consecutive teleportations on a qubit are often many blocks apart. So, the impact of a RELOCATE on future teleportations is only felt several blocks ahead.
Small lookahead windows, essential for low compilation complexity, miss these \textit{far-in-the-future} blocks.

Table~\ref{tab:tab-1} shows the average CNOT, block, and local-only block counts in between two consecutive RELOCATEs on a qubit in a 3$\times$3 DQC.
Consider the QAOA-FC benchmark with 57K CNOTs grouped into 17K blocks. 
On average, two consecutive RELOCATEs are separated by $\sim$90 blocks comprising $\sim$260 CNOTs, out of which $\sim$72 blocks have only local gates. Across all programs, consecutive RELOCATEs on a qubit are separated by 55-112 blocks on average. 
Small lookahead windows with only the next 10-15 blocks are thus, inadequate. Now one could argue that the overlapping qubit could be evicted from its location by an EPR release while scheduling the 17 (=89-72) blocks with non-local gates in between, rendering long-range lookahead ineffective.
However, note that the likelihood of this is extremely low because only $<$1 EPR release occurs on average between consecutive RELOCATEs and the likelihood of using the overlapping qubit is extremely low (given each chip has hundreds of qubits).

\begin{table}[htb]
\begin{center}
\caption{Average number of CNOTs, blocks, local-only blocks, and EPR releases between two consecutive RELOCATEs on a qubit in a $3\times3$ DQC for 240-qubit programs.} 
\setlength{\tabcolsep}{1mm} 
\renewcommand{\arraystretch}{1.}
\label{tab:tab-1}
\begin{small}
\begin{tabular}{ |c|c|c|c|c|}
\hline
Program & CNOTs & Blocks & Local-Only & EPR releases \\
\hline\hline

QAOA-FC & 261.6 & 89.6  & 72.1 & 0.12\\
\hline
QFT     & 137.2 & 55.3  & 33.2 & 0.02\\
\hline
QV      & 240.6 & 112.3 & 64.8 & 0.67\\
\hline
Shor    & 233.3 & 77.5  & 58.4 & 0.03\\
\hline
VQE     & 277.7 & 112.0 & 91.3 & 0.06\\
\hline
\end{tabular}
\end{small}
\end{center}
\end{table}

\begin{figure*}
    \centering
    \includegraphics[width=.991\linewidth]{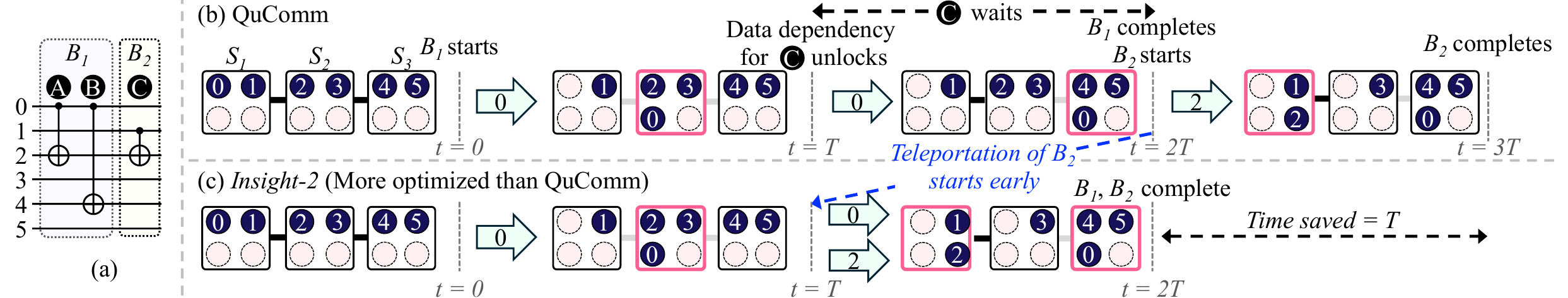}
    \caption{(a) An example circuit with blocks $B_1$ and $B_2$. 
    (b) Dependency for CNOT $C$ is resolved after executing CNOT $A$. But QuComm does not schedule $C$ until it reaches this gate in $B_2$ for scheduling. The teleportation required is also scheduled only when the compiler realizes that a relocation is necessary for scheduling $C$.
    (c) In contrast, \design{} relocates qubit $2$ and schedules CNOT $C$ \textit{as early as possible}. This improves RELOCATE and CNOT concurrency and reduces overall program execution time.}
    \label{fig:fig-3}
\end{figure*}

\begin{figure}[htpb]
    \centering
    \includegraphics[width=\linewidth]{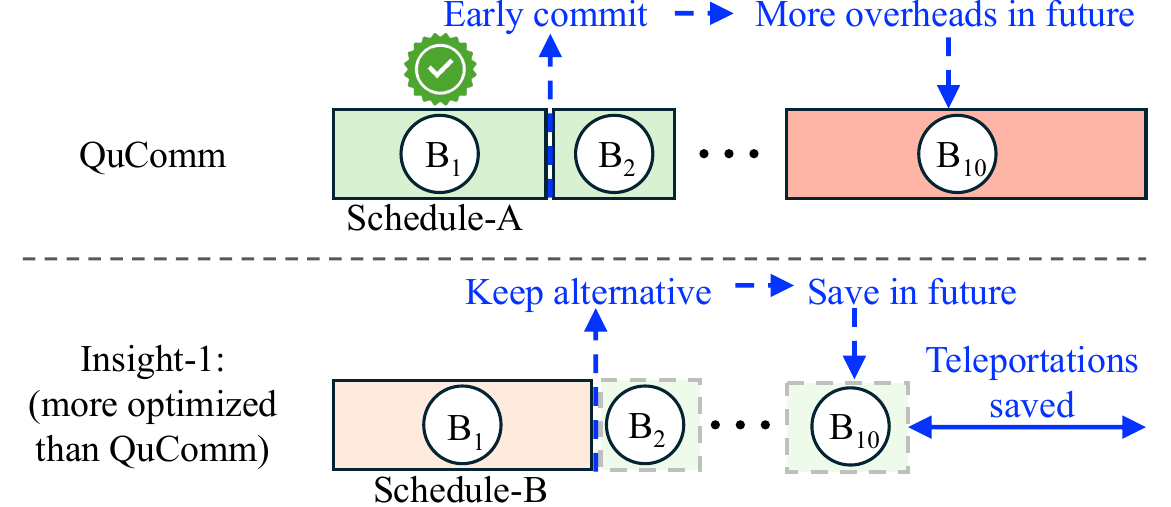}
    \caption{QuComm selects Schedule-A for $B_1$ as it costs fewer teleportations (illustrated by the box width here) than Schedule-B. But qubit relocations in Schedule-A eventually cost more teleportations in future block $B_{10}$. In contrast, \design{} also retains Schedule-B and schedules subsequent blocks, eventually arriving at a more optimized schedule for block $B_{10}$. This allows \design{} to select block schedules that are locally sub-optimal but globally better (like Schedule-B).}
    \label{fig:fig-4}
    \vspace{-5pt}
\end{figure}

The early commitment to a schedule also acts as a critical barrier because a schedule that is optimal for a block may eventually become \textit{globally sub-optimal} as future blocks are scheduled, as shown in Figure~\ref{fig:fig-4}. Unfortunately, existing compilers have no mechanisms to detect this sub-optimality and cannot pick an alternative schedule for a block scheduled in the past. For example, if schedule $A$ requires fewer teleportations than schedule $B$ for block $B_1$, QuComm selects $A$ before proceeding forward. However, it is possible that schedule $B$ would have resulted in significantly fewer teleportations for block $B_{10}$ and minimized teleportations overall. QuComm has no mechanisms to evaluate this because it discards $B$.

\vspace{0.05in}
\noindent \textbf{\textit{Insight-1: Practical utility-driven lookahead tailored to DQCs:}} DQC compilers must consider \textit{utility-driven lookahead}, where the lookahead window comprises only \textit{useful} future blocks rather than the next $F$ blocks. Moreover, to evaluate the cost of a relocation far ahead in future, compilers must avoid early commitment to a block schedule, and work with multiple schedules in parallel. Lastly, compilers must enable these optimizations without losing tractability. 

\subsection{Delayed Operation Scheduling}
Blocks comprise many gates that execute at different times, with some finishing sooner than others. The early operations of a block often resolve data dependencies for operations in future blocks. However, current compilers defer scheduling these gates until the compiler arrives at the corresponding future blocks for scheduling. Compilers also schedule teleportations \textit{on-demand}, 
which means that 
they determine the relocation paths only when they schedule a non-local CNOT and teleportations are required. These factors severely delay instruction scheduling, particularly long-latency teleportations, increasing program latencies. Our experiments across five benchmarks show that 51\% of teleportations are delayed on average, and each delayed teleportation waits for about 18\,ms on a 3$\times$3 DQC architecture, as shown in Table~\ref{tab:tab-2}.

\begin{table}[htb]
\begin{center}
\vspace{-5pt}
\caption{Percentage of delayed teleportations and their average wait time on a $3\times3$ DQC with 240-qubit programs.}
\label{tab:tab-2}
\setlength{\tabcolsep}{0.8mm}
\renewcommand{\arraystretch}{1.}
\begin{small}
\begin{tabular}{|l|c|c|c|c|c|}
\hline
                Metric  & QAOA-FC & QFT  & QV   & Shor & VQE  \\\hline\hline
Delayed teleportations (\%)   & 45.4    & 55.1 & 51.9 & 53.0 & 50.3 \\\hline
Avg. wait time (ms)        & 7.5     & 24.6 & 11.9 & 36.3 & 11.1 \\\hline
\end{tabular}
\end{small}
\vspace{-5pt}
\end{center}

\end{table}

We explain this with Figure~\ref{fig:fig-3}. 
Even though the data dependency for CNOT \circled{C} is resolved when CNOT \circled{A} executes ($t$=$T$), existing compilers wait to schedule it until block $B_1$ completes ($t$=$2T$), as shown in Figure~\ref{fig:fig-3}(b). Also, the RELOCATE for CNOT \circled{C} is only scheduled when the compiler arrives at that CNOT in $B_2$. In contrast, scheduling the RELOCATE and CNOT earlier, as shown in Figure~\ref{fig:fig-3}(c), reduces overall latency. 
The benefits compound in larger programs comprising hundreds to thousands of blocks because such early scheduling eventually leads to a cascade of operations to be scheduled sooner. 
For example, scheduling CNOT \circled{C} early unlocks data dependencies in subsequent blocks.

\vspace{0.05in}
\noindent \textbf{\textit{Insight-2: Decouple teleportations from CNOTs and schedule across block boundaries:}} DQC compilers must decouple teleportation scheduling from CNOT scheduling so that qubits are relocated ahead of time whenever feasible and EPR capacity is available. This allows the qubits to be available locally when the compiler is actually ready to schedule the corresponding CNOT. Compilers must also schedule gates as soon as possible, breaking the abstraction of the block boundary. This improves the instruction-level concurrency and reduces program runtimes. 

\begin{figure*}[htpb]    \centering\includegraphics[width=\linewidth]{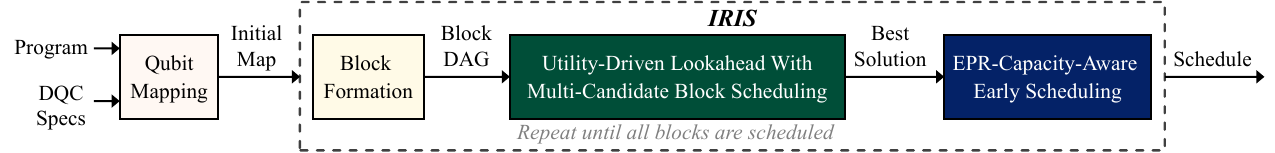}
    \caption{Overview of \design{}. It takes as input the program and initial qubit mapping and mainly focuses on instruction scheduling. This involves block formation (grouping operations into instruction blocks), an utility-driven lookahead-based scheduling involving multiple candidates, and optimizing teleportations via EPR-capacity-aware early scheduling.}
    \label{fig:fig-5}
\end{figure*}

\newpage
\section{Our Proposal: \design{}}

We propose \textit{\design{}} that reduces teleportation overheads during program execution by employing the insights described in the previous section.  
A DQC compiler involves two steps: mapping and scheduling. During mapping, it assigns a physical qubit to each program qubit. During scheduling, it starts with this initial layout to schedule instructions and updates the mapping based on qubit relocations during teleportations. \textit{\design{} mainly focuses on scheduling and its mapping is based on prior works.} Figure~\ref{fig:fig-5} gives an overview of \design{}.

\noindent \textbf{Pre-Processing:} Before compilation, physical qubits are split into compute (for program qubits) and communication qubits (for EPR pairs). By default, \design{} reserves 90\% of chip qubits for compute and 10\% for communication as non-local CNOTs occupy 3-8\% of operations in most programs. 

\subsection{Qubit Mapping}
\design{} maps qubits in three steps: (1)~partitions a program graph into $N$ sub-graphs (for DQC with $N$ chips), (2)~assigns each sub-graph to a chip, and (3)~maps each program qubit of a sub-graph to a compute qubit of the selected chip.

\vspace{0.05in}
\noindent \textit{\textbf{Step 1-- Program partitioning:}} This step minimizes non-local CNOTs based on the initial qubit layout. Partitioning methods can be broadly classified into two types: \textit{static} and \textit{dynamic}.
We consider both because \design{} is orthogonal to mapping and focuses on scheduling. 
Static methods, such as OEE~\cite{PARK1995899} and Min-Cut~\cite{wikipedia_minimum_cut}, partition the program at the beginning of compilation and fix the initial layout prior to scheduling. However, as scheduling proceeds and qubits relocate, some non-local CNOTs become local. Static methods cannot optimize RELOCATEs further based on dynamic qubit layout transitions as scheduling proceeds. So, \design{} also features state-of-the-art dynamic partitioning methods, such as WBCP~\cite{kaur2025optimized} and GCP~\cite{burt2024generalised}. They address this issue by dividing a program into multiple segments and selecting a separate qubit layout for each segment. As consecutive segment layouts may differ, these methods optimize for both the number of non-local CNOTs within each segment and RELOCATEs needed to transform one segment layout to the next.

By default, \design{} uses Min-Cut partitioning~\cite{wikipedia_minimum_cut}. While dynamic methods outperform static methods individually in general, our evaluations show that Min-Cut outperforms other mappers because the scheduler in \design{} outperforms the built-in scheduling integral to the dynamic schemes. We discuss these details in our evaluations.

\vspace{0.05in}
\noindent \textit{\textbf{Step 2-- Subgraph-to-chip mapping:}} Typically, DQCs have nearest-neighbor connectivity and sub-graphs with many non-local CNOTs between them must be placed on physically nearby chips to reduce collective RELOCATE costs~\cite{kaur2025optimized}. \design{} adopts the Integer Linear Programming (ILP) formulation proposed by Kaur et al~\cite{kaur2025optimized}  because it is a state-of-the-art.   

\vspace{0.05in}
\noindent \textit{\textbf{Step 3-- Subgraph-to-compute qubit mapping}:} This step assigns program qubits to compute qubits on a chip. It only affects the cost of on-chip gates depending on the qubit technology, 
and does not impact teleportation costs. \design{} applies a trivial mapping from program qubits within each chip to compute qubits and leaves optimization to each platform's monolithic compiler and is consistent with prior works~\cite{DQC-Wu2023QuComm,DQC-Wu2022AutoComm}.

\subsection{Block Formation}

\design{} initiates block formation by converting a program into a Directed Acyclic Graph (DAG) that captures data dependencies. The Block Formation Algorithm (BFA) starts with a current block $C$ set to the first gate $g_0$. BFA aggregates subsequent gates from the DAG that share one or more qubits with gates in $C$ (called overlapping qubits) into a candidate block $D$. BFA fuses $C$ and $D$ if the merged block, $C'$=$[C+D]$, can be executed on a chip with lower teleportation costs than executing $C$ and $D$ independently, and then updates $C$ to $C'$. If they cannot be merged due to insufficient EPR resources, BFA removes the last gate from $D$ and retries merging $C$ and $D$.
If $D$ is empty or the EPR capacity of the chip where $C$ is expected to execute has been reached, BFA saves $C$ as block $B_0$, sets $C$ to the next unassigned gate, and continues until all gates are assigned~to~blocks ($B_i$s).

The \textit{cost} ($c$) of executing a block on a chip is computed as the number of RELOCATEs ($n_{\textrm{relocate}}$) and Re-CNOTs ($n_{\textrm{remote}}$), using Equation~\eqref{eq:relocost}.
Re-CNOTs take longer than RELOCATEs. We account for this using the parameter $\alpha$. By default, we set $\alpha$=1.77 based on their latencies on neutral atom systems~\cite{Li2024Highrate,Bluvstein2024}. If there are multiple relocation paths to execute a block on a chip, BFA selects the one with minimum cost.
\begin{equation}
\label{eq:relocost}
     c = n_{\textrm{relocate}}+ \alpha \times n_{\textrm{remote}},
 \end{equation}

\begin{figure}[htbp]
    \centering    
    \includegraphics[width=\linewidth]{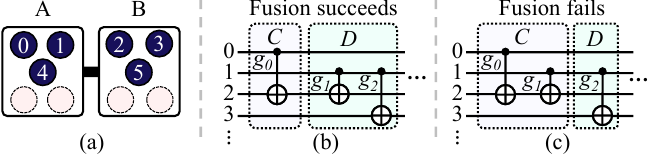}
    \caption{(a) A $1\times 2$ DQC. (b) BFA fuses blocks $C$ and $D$ because the cost of executing the combined block is lower than executing them separately and the required EPR capacity is available. (c) BFA does not fuse $C$ and $D$ because the execution cost of combined block exceeds the EPR capacity.} 
    \label{fig:fig-6}
\end{figure}

We explain BFA using Figure~\ref{fig:fig-6}.
 BFA starts with $C$=$[g_0]$ and $D$=$[g_1]$ because $g_1$ has qubit 2 overlapping with $C$.
 The cost of executing $C$ on either chip is 1 RELOCATE (qubit 0 from A to B or qubit 2 from B to A). Similarly, the cost of executing $D$ is 1 RELOCATE. Executing the fused block $C'$=$[g_0,g_1]$ costs 1 RELOCATE. Thus, $C$ is updated to $[g_0,g_1]$ and $D$ is set to $[g_2]$ (qubit 1 is the overlapping qubit). But the cost of the combined block $C'=[g_0, g_1, g_2]$, is now 2 RELOCATEs, exceeding EPR capacity. So, the fusion stops. 
 
\subsection{Utility-Driven Lookahead With Multi-Candidate Block Scheduling}

Next, starting with the initial qubit layout, \design{} proceeds to schedule each block. We propose \textit{\underline{U}tility-Driven Lookahead With \underline{M}ulti-Candidate Block \underline{S}cheduling (UMS)}. It employs two features: a \textit{utility-driven} lookahead window and the ability to retain multiple schedules in a solution tree.

\subsubsection{Utility-Driven Lookahead Window Selection} \hfill
\noindent 

The \textit{utility} 
of a \textit{future} block ($F$) determines the benefit of including it in the lookahead window of the current block ($C$), computed using Equation~\eqref{eq:utility}. If both blocks have overlapping qubit(s) and the compiler relocates at least one of them while scheduling $C$, there is a likelihood that the updated qubit positions will impact the relocations needed by $F$.
\begin{equation}
\label{eq:utility}
    \textrm{Utility}(F) = 1; \textrm{ if } Q \neq \emptyset, \textrm{~where }Q= \{q\mid q \in F \cap C\}
\end{equation}

\begin{figure}[htpb]
     \centering
     \includegraphics[width=.73\linewidth]{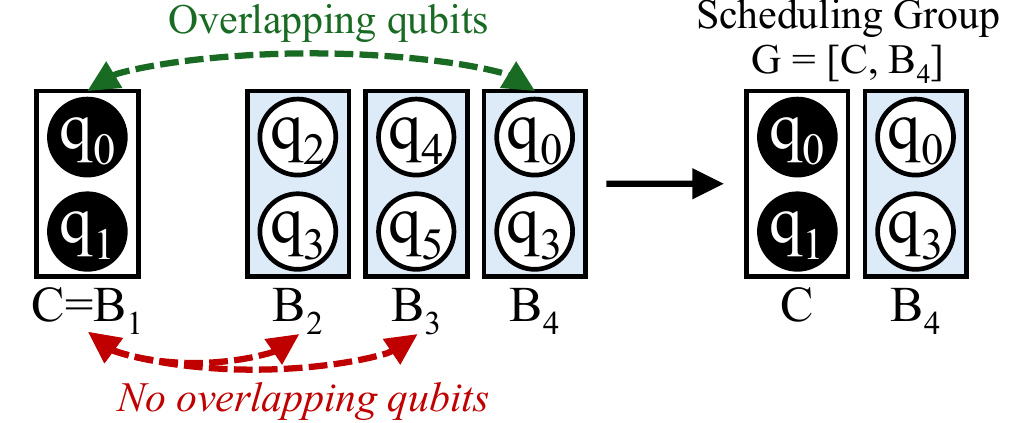}
     \caption{Illustration of utility-driven lookahead. Block $B_4$ has an overlapping qubit with current block $C$ and has utility in its lookahead but $B_2$ and $B_3$ do not. Scheduling group $G$ is the set of the current block and its lookahead window.}
     \label{fig:fig-7}
 \end{figure}

\design{} sets the next unscheduled block as the current block $C$ and prepares its lookahead window. It parses each subsequent block and checks for overlapping qubits. If there exists at least one overlapping qubit, the future block has \textit{utility} for $C$'s lookahead heuristic and is added to its lookahead window.
The process repeats until no blocks are left in the program or the lookahead window size is reached. Figure~\ref{fig:fig-7} illustrates this. As scheduling begins, $C$ is set to the first block $B_1$. The lookahead window does not include blocks $B_2$ and $B_3$ but includes $B_4$ because it has overlapping qubit ($q_0$) with $C$. 
The current block and its lookahead window form the scheduling group $G$ (i.e. $G{=[}C,B_4]$). By default, \design{} uses a lookahead window with 4 blocks, allowing it employ lookahead heuristics while keeping the complexity tractable (illustration shows only one block for brevity).

\subsubsection{Multi-Candidate Block Scheduling} \hfill
\label{subsubsec:apbs-group}

\noindent 
\design{} starts with an initially empty solution tree and maintain at most $w$ schedules at any time, where each branch or path corresponds to a schedule.
As scheduling proceeds, a layer of nodes is added for every non-local CNOT.
Each node comprises the schedule between two consecutive non-local CNOTs, qubit layout at the end of the corresponding schedule, and EPR capacity distributions.
\design{} schedules the current block $C$ in instruction order. 
Let $g$ be the next unscheduled gate. 
If $g$ is a local CNOT, it is scheduled and added to the newest layer of nodes. 
If $g$ is a non-local CNOT, \ums{} evaluates teleportation routes by considering (a)~which chips to execute $g$ and (b)~which qubits to relocate. 
For each combination of these, \ums{} selects between RELOCATEs and Re-CNOTs depending on what is cheaper, identifies relocation paths, and checks if EPR capacity is available. 
If the EPR capacity of a chip along the relocation path is exhausted, \ums{} frees EPR capacity using the \textit{EPR release} and prioritizes the eviction of qubits whose relocation benefits future gates in group $G$. 
If no such qubit is found, \ums{} evicts an external qubit. 
\ums{} minimizes the cost of scheduling $g$ using Equation~\eqref{eq:ctotal}, computed as the sum of its teleportation costs ($C_g$), any EPR releases ($C_{\textrm{EPR}}$), and its impact ($C_R$) on the set of remaining non-local CNOTs ($R$) in the scheduling group $G$. 
\begin{equation}
    \label{eq:ctotal}
   \textstyle{C_{\text{total}}(g) = C_g + C_{\textrm{EPR}} + C_R}.
\end{equation}

 Evaluating $C_R$ is non-trivial because teleportation costs of a future gate cannot be determined without actually scheduling it starting with qubit positions at the end of its preceding gate. So, \ums{} only \textit{estimates} $C_R$ based on current qubit positions, using Equation~\eqref{eq:future}, where $C_i$ is the teleportation cost of the $i^{th}$ non-local gate in $R$, $\beta$ is a decaying factor based on distance ($d$) between the blocks comprising gates $g$ and $i$ which allows \ums{} to prioritize future gates based on their distance from current gate. As discussed earlier, EPR releases do not occur frequently and so, the qubit being relocated for $g$ typically stays at the same location when gates in $R$ are scheduled. As a result, our estimated $C_R$ closely matches the actual $C_R$. 
 By default, we set $\beta = 0.871$ so that a block that is $10$ blocks ahead in the future contributes a weight $0.25$. 

\begin{equation}
    \label{eq:future}
    \textstyle{C_{\textit{R}} = \sum_{i\in R} \beta^{d(i)} \times C_i.}
\end{equation}

\begin{figure}[htpb]
     \centering
     \includegraphics[width=\columnwidth]{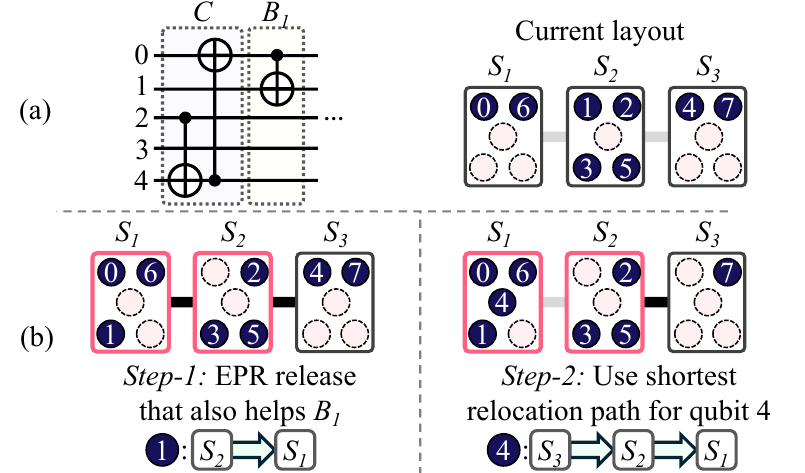}
     \caption{(a) A scheduling group comprising the current block $C$ and its utility-driven lookahead window $B_1$ being scheduled assuming a current qubit layout. (b) \ums{} frees EPR resources on chip $S_2$ by relocating qubit 1 to $S_1$. This relocation also helps in reducing the teleportation costs of the CNOT in block $B_1$. To relocate qubit 4, \ums{} uses the shortest relocation path, allowing it to minimize  teleportation costs.}\label{fig:fig-8}
\end{figure}

We explain scheduling using Figure~\ref{fig:fig-8}. Assume a chip has an initial EPR capacity of two and qubits ($0,1$), ($2,3$), and ($4,5$) are mapped to chips $S_1$, $S_2$, and $S_3$, respectively. Although $C$ and $B_1$ share an overlapping qubit, block formation does not merge them to avoid exceeding the EPR capacity. Consider the scheduling group $G{=}[C, B_1]$ with $C$ as the current block and the qubit layout in Figure~\ref{fig:fig-8}. To schedule $C$, qubit $4$ must relocate from $S_3$ which requires an EPR release on $S_2$. For this, \ums{} evicts qubit $1$ to $S_1$ that makes the CNOT in $B_1$ local. \ums{} executes the first CNOT in $C$ between qubits $2$ and $4$ on $S_2$ along the relocation path of qubit $4$.

\begin{figure}[htpb]
    \centering
    \includegraphics[width=.95\linewidth]{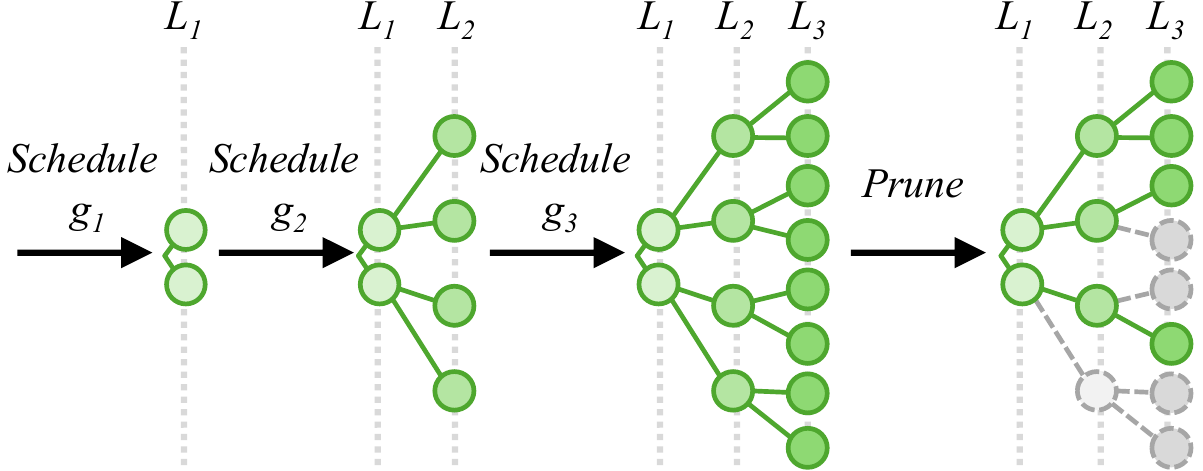}
    \caption{Nodes are added to the solution tree for each non-local gate scheduled depending on relocation paths (here, each non-local CNOT adds two paths). Once the tree exceeds $w$ solutions, it is pruned to only retain the top-$w$ candidates.}
    \label{fig:fig-9}
\end{figure}

For each non-local gate $g$, \ums{} starts with qubit layout in each node in the last layer (say $L_i$) of the tree and considers $p$ relocation paths for $g$, where $p$ is the number of chips comprising qubits of remaining operations in $G$. For example, if the qubits involved in the operations remaining in $G$ after $g$ are on two chips, \ums{} considers $p{=}2$ relocation paths for $g$. Once the paths are determined, nodes are added to the next layer ($L_{i+1}$) of the tree. When \ums{} completes evaluation for all nodes in layer $L_i$, the tree is pruned to retain only the top-$w$ solutions. Figure~\ref{fig:fig-9} illustrates this process. When gate $g_1$ is scheduled, \ums{} considers two relocation paths, adding two nodes to the tree. The next non-local gate $g_2$ schedules starting from each $L_1$ node and considers two relocation paths per node. Scheduling proceeds to gate $g_3$ starting with the $L_2$ nodes. Assuming each node adds two relocation paths for $g_3$, the tree now comprises eight paths which exceeds the maximum capacity ($w{=}4$) and the tree is pruned to only retain the top four candidates. Once all gates in the current block $C$ are scheduled, \design{} sets $C$ to the next unscheduled block. The process continues until all the blocks in the program are scheduled. After this, \design{} returns the schedule from the tree with the fewest teleportations.

\subsection{EPR-Capacity-Aware Early Scheduling (\ees{})}

\design{} executes future gates and teleportations early \textit{without waiting} for the current block to complete. Our proposed method, \textit{\underline{E}PR-Capacity-Aware \underline{E}arly \underline{S}cheduling (\ees{})} enables two capabilities: (1) early scheduling of a RELOCATE required for a future CNOT, if the qubit involved in the RELOCATE does not participate in any other operations, even if the data dependencies for the CNOT are not yet resolved, which allows us to decouple teleportation from CNOT scheduling and (2) early scheduling of future CNOTs and any RELOCATEs they need if data dependencies are resolved.

But such early execution is non-trivial because dependent operations are often blocks apart and executing RELOCATEs early may alter EPR capacities and qubit positions that interfere with scheduling intermediate operations. For example, if a gate in block $B_1$ resolves the dependency for a gate in $B_5$, scheduling the $B_5$ gate early may alter qubit positions and EPR resource distributions such that scheduling pending operations in $B_1$ to $B_4$ costs extra relocations. These costs are unknown until the gates are scheduled. 
Also, tracking dependencies is hard as they span several blocks (say $B_1$$ \rightarrow $$B_5$$ \rightarrow $$B_{10}$). \ees{} overcomes these challenges using a two-pass approach. 

First, \ees{} schedules the program using \ums{}, finds a set of all instructions $E$ that may be executed earlier than their current timelines by tracing their dependency chains, and tracks qubit positions and EPR capacities after each instruction. Next, for each instruction $e$ in $E$, \ees{} checks its current execution timestamp ($T_f$) and the earliest timestamp ($T_e$) it can execute. Then, \ees{} checks if $e$ can be executed a cycle earlier than $T_f$ (at $T_i$=$ T_f $$-$$\tau$) without extra relocations. Here, $\tau$ is the execution time of the previous instruction. 
If $e$ needs extra relocations to execute at $T_i$, \ees{} removes it from $E$ and schedules it at its current timeline. 
Otherwise, it updates the schedule by shifting $e$ earlier to $T_i$.
If $e$ is a RELOCATE, \ees{} executes it at $T_i$ only if it does not consume the full EPR capacity of the corresponding chip. 
This ensures subsequent instructions do not need EPR releases. 
The process continues until all timestamps up to $T_e$ are evaluated. 
The steps are repeated for all instructions in $E$ before moving forward. 
After \ees{} processes all instructions, it returns the final schedule.

\begin{figure}[htpb]
    \centering
    \includegraphics[width=\linewidth]{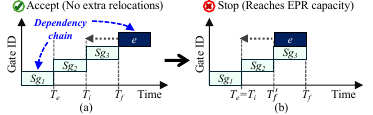}
    \caption{(a) \ees{} finds the earliest timestamp $T_e$ at which a future instruction $e$ can execute. It shifts $e$ from its timestamp $T_f$ to an earlier one, $T_i$, if it does not add relocations. (b) \ees{} stops shifting $e$ earlier than $T_i$ if a chip reaches EPR capacity.}
    \label{fig:fig-10}
\end{figure}

Figure~\ref{fig:fig-10} illustrates \ees{}.
Here, $e$ is an early executable instruction whose dependency is resolved after $Sg_0$ executes. \ees{} shifts the execution of $e$ from $T_f$ to $T_i$ as it does not incur extra relocations.
But, \ees{} does not shift the execution to $T_e$ as the EPR capacity of the corresponding chip is reached. 

\vspace{0.02in}
We summarize the \design{} algorithm in Appendix~\ref{appendix:algorithm}.

\section{Evaluation Methodology}
This section describes our evaluation methodology.

\subsection{Benchmarks}

\label{subsec:benchmarks}

We evaluate \design{} using a wide array of benchmarks, ranging from arithmetic circuits to system-level benchmarking programs and variational quantum algorithms~\cite{bench-bv,bench-ising,bench-qaoa,bench-qft,bench-qv,bench-vqe}. 
Program sizes (up to 86.4K CNOTs) vary by DQC size, and far exceed those used in prior works. 
Table~\ref{tab:tab-3} summarizes them.

\begin{table}[htpb]
\renewcommand{\arraystretch}{1.2}
\setlength{\tabcolsep}{2pt}
\caption{CNOT counts in benchmarks for $2\times2$, $2\times3$, and $3\times3$ DQCs (with 120, 180, and 240 program qubits, respectively).}
\begin{small}
\begin{tabular}{|l||r|r|r|r|r|r|r|r|}
    \hline
    \textbf{DQC} &
    \textbf{BV} &
    \makecell{\textbf{QAOA}\\ \textbf{3reg}} &
    \textbf{QuGAN} &
    \textbf{VQE} &
    \textbf{QFT} &
    \textbf{Shor} &
    \makecell{\textbf{QAOA}\\ \textbf{FC}} &
    \textbf{QV} \\
    \hline\hline
    $2\times2$ & 119 & 360 & 706  & 7.2K & 7.3K & 13.6K & 14.3K & 21.6K \\
        \hline
    $2\times3$ & 179 & 540 & 1.1K & 16.2K & 16.4K & 31.1K & 32.2K & 48.6K \\
        \hline
    $3\times3$ & 239 & 720 & 1.4K & 28.8K & 29.0K & 55.9K & 57.4K & 86.4K \\
    \hline
\end{tabular}
\end{small}
\label{tab:tab-3}

\end{table}

\subsection{Baseline}\label{subsec:baseline}
\noindent \textbf{Mapping:} We employ WBCP~\cite{kaur2025optimized}, GCP-E~\cite{burt2024generalised}, static OEE~\cite{PARK1995899} and Min-Cut~\cite{wikipedia_minimum_cut} for circuit partitioning and Kaur's ILP method~\cite{kaur2025optimized} for mapping sub-graphs onto chips, which represents the state-of-the-art, enabling a competitive baseline.  

\vspace{0.05in}
\noindent \textbf{Program Scheduling:} We compare against QuComm~\cite{DQC-Wu2023QuComm}, the state-of-the-art scheduler for DQCs. As its software is not open sourced, we implement and reproduce it faithfully. 


\subsection{Hardware Architectures}

\noindent \textbf{Architectures:} We consider $2$$\times$$2$, $2$$\times$$3$, and $3$$\times$$3$ DQCs to match industry roadmaps~\cite{ionq_roadmap_2024,pasqal_roadmap_2025,PsiQuantum2025Photonic}. We consider 100--500 qubits systems and varying chip sizes, which far exceed the 300-qubits maximum used in prior works~\cite{Bluvstein2024,lin2025zac,Ruan2025ASPLOS_Powermove,DQC-Wu2023QuComm,DQC-Wu2022AutoComm}.

\vspace{0.05in}
\noindent \textbf{Error Model:} We assume  neutral-atoms~\cite{Bluvstein2024} because they have high photonic link rates and fidelity~\cite{Li2024Highrate}. We consider dual-atom architectures where Rubidium ($^{87}$Rb) and Ytterbium ($^{171}$Yb) atoms are used for compute and communication qubits respectively. 
This heterogeneity allows frequent measurements and teleportation-related operations on communication qubits while minimizing crosstalk with compute qubits~\cite{zhang2025dual,Anand2024DualSpeciesRydbergArray,Shen2024DualSpeciesRbYbAtomArray}. We summarize the parameters in Table~\ref{tab:tab-4}.

\begin{table}[htp]
    \centering
    \small
    \renewcommand{\arraystretch}{1.}
    \setlength{\tabcolsep}{5pt}
    \caption{Operational latencies in neutral atom DQCs~\cite{Bluvstein2024,Li2024Highrate}}

    \begin{tabular}{|l|r||l|r|}
        \hline
        \textbf{Chip Params.} & \textbf{Time} & 
        \textbf{Interconnect Params.} & \textbf{Time} \\
        \hline
        \hline
        1Q gate & 52\,$\upmu$s & Measurement & 1\,ms \\
        \hline
        2Q gate & 360\,ns & EPR Pair Generation & 259\,$\upmu$s \\
        \hline
        Atom transfer & 15\,$\upmu$s & RELOCATE & 1.3\,ms  \\
        \hline
        Atom movement & 300\,$\upmu$s 
        & Re-CNOT & 2.3\,ms \\
        \hline
    \end{tabular}

    \label{tab:tab-4}
\end{table}

\begin{table*}
\centering
\caption{Effective teleportation count ($T_{\text{eff}}$) and latency ($L$ in seconds) comparison using different mappers on $2\times2$ DQC.}
\renewcommand{\arraystretch}{1.01}
\setlength{\tabcolsep}{4.5pt}
    \label{tab:main_table_2x2}
        \begin{small}
    \begin{tabular}{|c||c|c|c|c||c|c|c|c||c|c|c|c||c|c|c|c|}
    \hline
    \multirow{3}{*}{Program}    & \multicolumn{4}{c||}{Mapper-1: GCP-E} & \multicolumn{4}{c||}{Mapper-2: sOEE} & \multicolumn{4}{c||}{Mapper-3: WBCP} &
    \multicolumn{4}{c|}{Mapper-4: Min-Cut \textbf{\textit{(Best)}}} \\
    \cline{2-17}
    \cline{2-17}
    & \multicolumn{2}{c|}{QuComm} & \multicolumn{2}{c||}{\design{}} & \multicolumn{2}{c|}{QuComm} & \multicolumn{2}{c||}{\design{}} & \multicolumn{2}{c|}{QuComm} & \multicolumn{2}{c||}{\design{}} &
    \multicolumn{2}{c|}{QuComm} & \multicolumn{2}{c|}{\design{}} \\
    \cline{2-17}
    \cline{2-17}
        & $T_{\mathrm{eff}}$ & $L$ & $T_{\mathrm{eff}}$ & $L$  & $T_{\mathrm{eff}}$ & $L$ & $T_{\mathrm{eff}}$ & $L$  & $T_{\mathrm{eff}}$ & $L$ & $T_{\mathrm{eff}}$ & $L$ & $T_{\mathrm{eff}}$ & $L$ & $T_{\mathrm{eff}}$ & $L$ \\
    \hline
    \hline
BV & 95 & 0.16 & 85 & 0.10 & 95 & 0.16 & 83 & 0.10 & 27 & 0.07 & 21 & 0.06 & 73 & 0.13 & 60 & 0.09\\\hline
QAOA-3reg & 160 & 0.17 & 139 & 0.07 & 105 & 0.18 & 83 & 0.07 & 161 & 0.20 & 123 & 0.07 & 77 & 0.16 & 57 & 0.07\\\hline
QAOA-FC & 1189 & 1.97 & 682 & 0.96 & 1551 & 2.09 & 737 & 0.95 & 987 & 2.35 & 734 & 1.03 & 906 & 2.11 & 542 & 0.90\\\hline
QFT & 1126 & 1.41 & 944 & 0.78 & 1212 & 1.58 & 930 & 0.78 & 800 & 1.54 & 697 & 0.81 & 988 & 1.53 & 656 & 0.84\\\hline
QuGAN & 283 & 0.54 & 188 & 0.27 & 93 & 0.34 & 80 & 0.20 & 141 & 0.39 & 112 & 0.24 & 36 & 0.26 & 18 & 0.18\\\hline
QV & 7182 & 3.35 & 5979 & 1.68 & 7192 & 3.52 & 6060 & 1.70 & 7244 & 3.36 & 6039 & 1.62 & 7238 & 3.43 & 5965 & 1.67\\\hline
Shor & 1430 & 1.95 & 870 & 0.90 & 1824 & 2.34 & 660 & 0.97 & 1179 & 2.22 & 602 & 0.92 & 1751 & 2.39 & 644 & 0.92\\\hline
VQE & 912 & 1.12 & 665 & 0.58 & 910 & 1.24 & 765 & 0.54 & 622 & 1.27 & 536 & 0.58 & 817 & 1.17 & 629 & 0.59\\\hline\hline
\textbf{Relative} & {1} & {1} & {\textbf{0.74}} & {\textbf{0.50}} & {0.89} & {1.01} & {\textbf{0.62}} & {\textbf{0.49}} & {0.68} & {0.94} & {\textbf{0.52}} & {\textbf{0.47}} & {0.66} & {0.93} & {\textbf{0.42}} & {\textbf{0.47}}\\
\hline
    \end{tabular}
    \end{small}
\end{table*}

\vspace{0.05in}
\noindent \textbf{Assumptions on EPR Pair Generation:} EPR pair generation is probabilistic and error-prone. Producing high-fidelity EPR pairs requires \textit{distillation} performed by repeating a Bell-State Measurement (BSM) process~\cite{Li2024Highrate,Pan2001Purification,Bennett1996Purification}. 
Each attempt takes $1\mu s$ and produces a successful entanglement with probability $p$ = $1/8$~\cite{Li2024Highrate}.
After $n$ attempts, this probability increases to $\textsc{P}$ = $1-(1-p)^n$.
We set $n=52$ to achieve $P=99.9\%$, requiring $52\,\mu s$. The resulting Bell pairs have a fidelity of ${\sim}0.999$~\cite{Li2024Highrate}.
EPR pair generation is multi-step process that includes loading the communication qubit from the chip to an optical cavity ($100 \mu s$), optical pumping and application of a $\pi/2$ pulse ($1.1 \mu s$), repeated BSM ($52 \mu s$), cavity de-pumping ($6 \mu s$), unloading the communication qubit from the cavity to the chip ($100 \mu s$), which takes $259 \mu s$. However, this latency can be hidden with program operations by initiating the process ahead of time. So, if an EPR pair is required at time $T$, we assume that EPR pair generation starts at time $t$$=$$T$$-$259$ \mu s$. 

\subsection{Figures of Merit}

\noindent \textbf{Effective Teleportation Count ($T_{\textrm{eff}}$):} We compute effective teleportation count using Equation~\eqref{eq:effteleportation}, where $N_\textrm{{RELOCATE}}$ and $N_{\textrm{Re-CNOT}}$ are the number of RELOCATEs and Re-CNOTs in the schedule. As Re-CNOTs take $1.77\times$ longer than RELOCATEs, we set $\alpha$=1.77. It can be adjusted if gate characteristics change or \design{} is adopted in other technologies. 
\begin{equation}
\label{eq:effteleportation}
     T_{\textrm{eff}} = N_{\textrm{RELOCATE}}+ \alpha \times N_{\textrm{Re-CNOT}},
 \end{equation}

\noindent \textbf{Latency (L):} We also measure the end-to-end latency of a program schedule, similar to prior works~\cite{DQC-Wu2022AutoComm,DQC-Wu2023QuComm}.

\vspace{0.5em}
 \textit{Lower teleportation counts ($T_{\mathrm{eff}}$) and  latency (L) are better.}

\subsection{Experimental Setup}

We conduct experiments on a  server equipped with 160-core ARM Neoverse-N1, 512~GB RAM, and 2~TB SSD. Our software runs 
in Python~3.9 environment including \texttt{qiskit}~\cite{qiskit} to parse circuits, \texttt{networkx}~\cite{networkx}, \texttt{pymetis}~\cite{pymetis}, and \texttt{pulp}~\cite{pulp} for the circuit partitioning and ILP-based chip selection.


\section{Results}
\subsection{Effective Teleportation Count and Latency}
Table~\ref{tab:main_table_2x2} shows the effective teleportation count ($T_{\mathrm{eff}}$) and latency of \design{} and QuComm for different mappers on a 2$\times$2 DQC. \design{} consistently outperforms QuComm for all mappers, showing that it is generalizable. By default, \design{} uses the Min-Cut mapper, because it outperforms the rest and dynamic mappers (WBCP and GCP-E) have their own lookahead-driven teleportation scheduling that conflicts with \design{}, eventually degrading performance. 
\design{} reduces $T_{\mathrm{eff}}$ by 36\% on average and by up to 63\%; and reduces latency by 0.51$\times$ on average and by up to 0.39$\times$ compared to QuComm. Table~\ref{tab:main_table_others} compares QuComm and \design{} on $2$$\times$$3$ and $3$$\times$$3$ DQCs. 
\design{} reduces $T_{\mathrm{eff}}$ by 35\% on average (up to 65\%) on the $2$$\times$$3$ DQC and by 32\% on average (up to 56\%) on the $3$$\times$$3$ DQC.
It reduces latency to $0.50\times$ on average on the $2$$\times$$3$ DQC and to $0.48\times$ on average on the $3$$\times$$3$ DQC, and up to $0.35\times$ and $0.36\times$ on $2$$\times$$3$ and $3$$\times$$3$ DQCs respectively.

\begin{table}[htpb]
\centering
\caption{Effective teleportation count ($T_{\text{eff}}$) and latency ($L$ in seconds) on 2$\times$3 and 3$\times$3 DQCs using Min-Cut mapping.}
\renewcommand{\arraystretch}{1.01}
\setlength{\tabcolsep}{2.3pt}
    \label{tab:main_table_others}
    \begin{small}
    \begin{tabular}{|c||c|c|c|c||c|c|c|c|}
    \hline
    \multirow{3}{*}{Program}    & \multicolumn{4}{c||}{2$\times$3} & \multicolumn{4}{c|}{3$\times$3} \\
    \cline{2-9}
    \cline{2-9}
    & \multicolumn{2}{c|}{QuComm} & \multicolumn{2}{c||}{\design{}} & \multicolumn{2}{c|}{QuComm} & \multicolumn{2}{c|}{\design{}} \\
    \cline{2-9}
    \cline{2-9}
        & $T_\text{eff}$ & $L$ & $T_\text{eff}$ & $L$  & $T_\text{eff}$ & $L$ & $T_\text{eff}$ & $L$  \\
    \hline    
    \hline

BV & 164 & 0.27 & 148 & 0.18 & 316 & 0.49 & 262 & 0.28\\\hline
QAOA-3reg & 201 & 0.26 & 153 & 0.11 & 378 & 0.44 & 289 & 0.16\\\hline
QAOA-FC & 3357 & 4.34 & 1843 & 1.72 & 7640 & 6.79 & 4802 & 2.94\\\hline
QFT & 3974 & 3.37 & 1980 & 1.59 & 7298 & 4.40 & 4022 & 2.52\\\hline
QuGAN & 89 & 0.43 & 61 & 0.32 & 186 & 0.66 & 106 & 0.41\\\hline
QV & 22496 & 6.28 & 19473 & 3.14 & 51740 & 9.81 & 44901 & 4.70\\\hline
Shor & 4642 & 4.90 & 1608 & 1.69 & 7861 & 6.62 & 3471 & 2.59\\\hline
VQE & 1998 & 2.12 & 1741 & 1.20 & 4367 & 3.42 & 4279 & 1.65\\\hline\hline
\textbf{Relative} & 1.00 & 1.00 & \textbf{0.65} & \textbf{0.50} & 1.00 & 1.00 & \textbf{0.68} & \textbf{0.48}\\\hline
    \end{tabular}
    \end{small}
\end{table}


\subsection{Impact of \ums{}}
\begin{figure}[htpb]
    \centering      

        \includegraphics[width=\columnwidth]{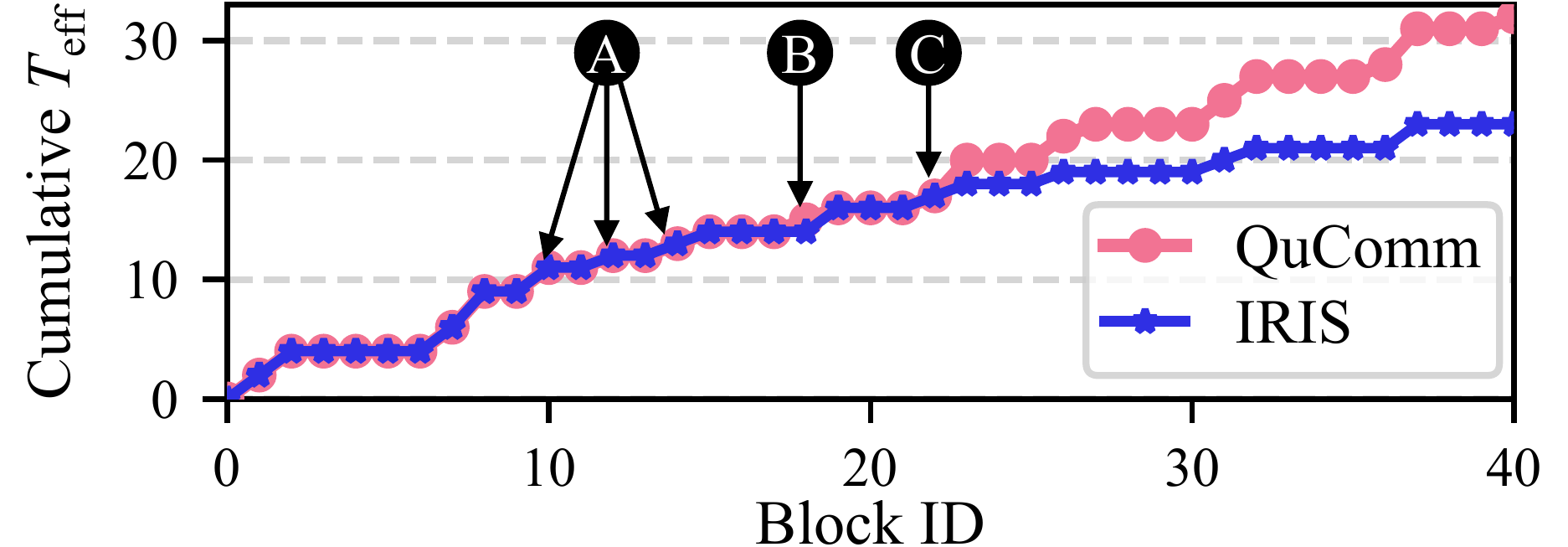}  
    \caption{Cumulative teleportation count as scheduling proceeds in QuComm and IRIS for a 240-qubit Shor's algorithm. Utility-driven lookahead with multi-candidate block scheduling in IRIS reduces total teleportation costs.}\label{fig:ums_blockid_teff}
\end{figure}

Figure~\ref{fig:ums_blockid_teff} compares the cumulative $T_{\text{eff}}$ as scheduling proceeds for a 240-qubit Shor's algorithm on a 3$\times$3 DQC. Blocks marked by \circled{A} denote cases where \design{} selects a different chip compared to QuComm to execute the block. These selections eventually reduce teleportations in the subsequent block \circled{B}. A different scheduling for block \circled{C} reduces teleportations for all blocks beyond Block-23.

\subsection{Accuracy of Future $T_{\textrm{eff}}$ Estimation}
\begin{figure}[b]
    \centering      
\includegraphics[width=\columnwidth]{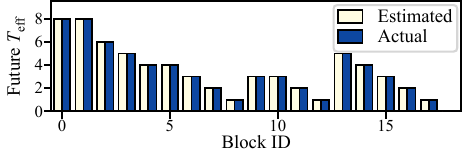}
\caption{Estimated teleportation counts during scheduling are identical to actual costs during execution (data is obtained from a 32-qubit QAOA 3-regular program on $2\times2$ DQC).}
    \label{fig:ums_blockid_future_teff}

\end{figure}

Recollect that IRIS estimates the costs of teleportations of a future block while scheduling.
Figure~\ref{fig:ums_blockid_future_teff} shows that the estimated and actual teleportation are identical costs for all blocks in a 32-qubit QAOA 3-regular program on a $2\times2$ DQC system. This is because relocated qubits often do not undergo teleportations, operations, or EPR-releases in between.

\subsection{Impact of \ees{}}
\begin{figure}[htpb]
    \centering
\includegraphics[width=\columnwidth]{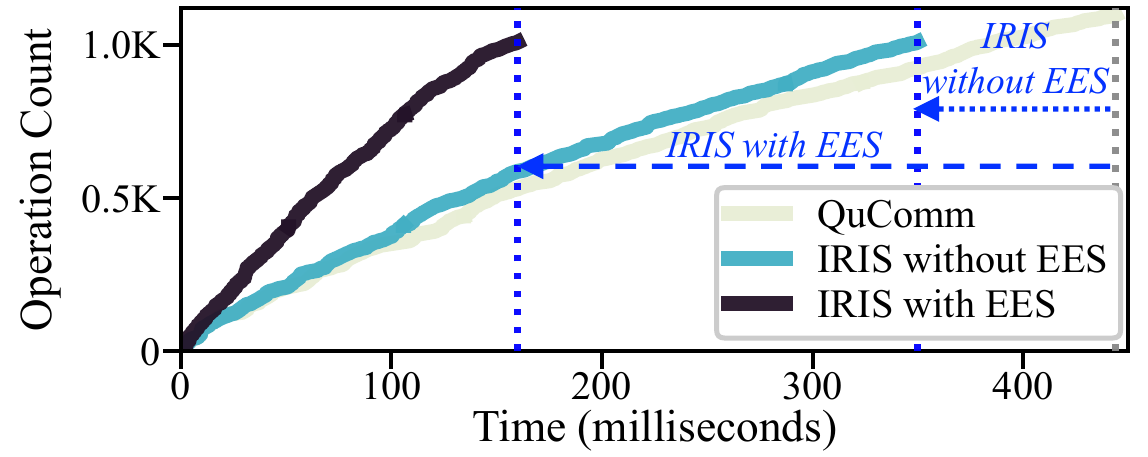}
    \caption{Schedules of a 240-qubit QAOA 3-regular program. \design{} requires fewer RELOCATEs and achieves higher RELOCATE concurrency, leading to lower program latency.}
    \label{fig:results_ees}

\end{figure}

Figure~\ref{fig:results_ees} shows the schedule of a 240-qubit QAOA-3reg program on a $3\times3$ DQC. The schedule from QuComm takes {443}\,ms and has a
RELOCATE concurrency of only {0.85}~RELOCATE/ms.
\design{} (without \ees{}) reduces the latency to 348\,ms with a RELOCATE concurrency of {0.83}~RELOCATE/ms. With \ees{}, \design{} reduces latency to
{160}\,ms and achieves an even higher RELOCATE concurrency of {1.81}~RELOCATE/ms. 
Also, \design{} (without \ees{}) improves latency by 1.2$\times$ on average and up to 1.6$\times$ compared to QuComm. With \ees{}, the latency reduces to 2.0$\times$ on average and by up to 2.9$\times$. 

\subsection{Impact of Each Optimization}

\begin{figure}[htpb]
\centering
\includegraphics[width=\columnwidth]{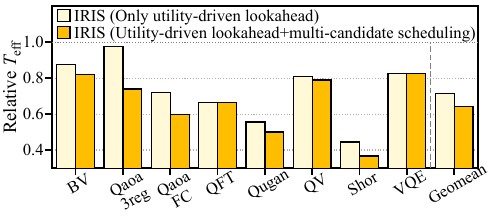}
\includegraphics[width=\columnwidth]{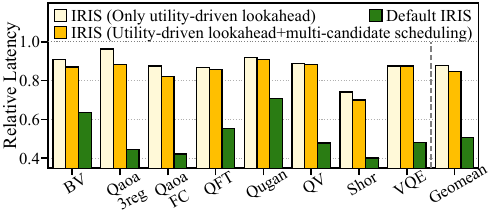}\\[0pt]
\caption{Impact of each optimization in IRIS on a $2$$\times$$2$ DQC for various programs. Values are normalized to QuComm.  
}
\label{fig:each_optimization}

\end{figure}

Figure~\ref{fig:each_optimization}
shows the efficacy of each optimization in \design{}.

Only utility-driven lookahead reduces $T_{\textrm{eff}}$ and latency by 1.40$\times$ and 1.14$\times$ on average, respectively. Introducing multi-candidate scheduling on top further reduces $T_{\textrm{eff}}$ and latency by 1.57$\times$ and 1.18$\times$. Finally, introducing \ees{} reduces latency by 1.98$\times$ on average compared with QuComm. 
Note that both utility-driven lookahead and multi-candidate scheduling reduce both $T_{\textrm{eff}}$ and latency. \ees{} does not reduce $T_{\textrm{eff}}$, but only improves latency by making a schedule more compact.

\begin{figure}[htpb]
\centering
\includegraphics[width=.99\columnwidth]{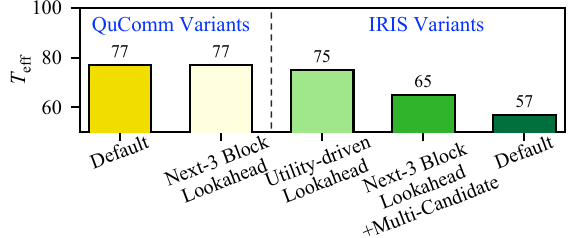}
\caption{Effective teleportation count of a 120-qubit QAOA 3-regular graph with different QuComm and \design{} variants.}
\label{fig:ums_variants}
\end{figure}

We also compare the default \design{} against multiple variants of QuComm and \design{}, as shown in
Figure~\ref{fig:ums_variants}, for a 120-qubit QAOA 3-regular program on a $2\times2$ DQC. 
For QuComm, we consider next-3 blocks for lookahead, besides the default. For IRIS, we consider three variants: (a)-with one candidate and utility-driven lookahead, (b)-next-3 blocks for lookahead with multi-candidate scheduling, and (c)-default (both utility-driven lookahead and multi-candidate scheduling).
QuComm with next-3 blocks lookahead is ineffective. Similarly, multi-candidate scheduling outperforms single candidate variants.

\subsection{Impact of Scaling System Sizes}
\begin{figure}[htpb]
\centering    \includegraphics[width=\linewidth]{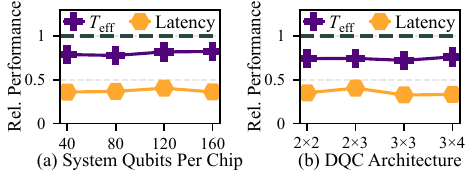}
    \caption{Impact of scaling on IRIS performance using QAOA 3-regular programs relative to QuComm. (a) Scaling the chip size
    on a 2$\times$2 DQC. (b) Scaling the number of chips.
    }
    \label{fig:result_scaling}
\end{figure}

Figure~\ref{fig:result_scaling}(a) evaluates \design{} across four chip sizes on a 2$\times$2 DQC using QAOA 3-regular programs. The number of qubits per chip ranges from 40 to 160 and program sizes range from 120 to 500 qubits. \design{} reduces $T_{\text{eff}}$ by 18--22\% and latency by 60--64\%. While \design{} consistently outperforms QuComm for all configurations, we cannot conclude on any scaling trends for $T_{\text{eff}}$ because the programs differ in sizes. The latency improves with increasing chip (and program) sizes because \design{} has more opportunities to optimize the schedules. Figure~\ref{fig:result_scaling}(b) shows the impact of increasing the number of chips. 
\design{} consistently lowers $T_{\text{eff}}$ and latency. 

\subsection{Sensitivity to Design Parameters}
\begin{figure}[htpb]

    \centering
\includegraphics[width=\linewidth]{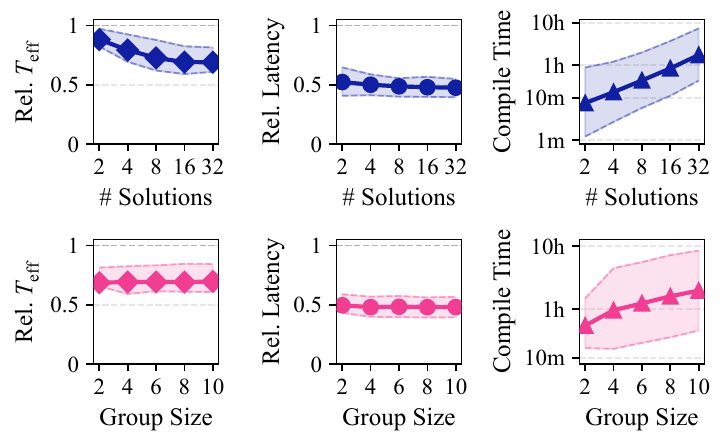}
    \caption{Impact of $w$ and $|G|$ on \ums{}. Values are relative to QuComm based on average from over 24 benchmarks.}
    \label{fig:results_beamwidth}
\end{figure}

Figure~\ref{fig:results_beamwidth} shows the impact of increasing the number of solutions in the tree ($w$) and scheduling group sizes ($|G|$) on the performance and runtime of \design{}. While increasing $w$ initially improves the performance, the returns diminish beyond $w{=}16$. Increasing group sizes does not significantly improve performance beyond $|G|{=}4$. Increasing both leads to increased compilation time which is expected due to increased search space. By default, \design{} uses ($w$,$|G|$)=($16$,$4$).

\subsection{Sensitivity to Hardware Parameters}
\begin{figure}[htpb]
\centering
\includegraphics[width=\columnwidth]{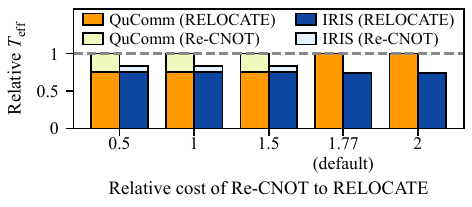}
\caption{IRIS performance relative to QuComm for varying Re-CNOT to RELOCATE cost ratios.}
\label{fig:alpha}
\end{figure}

\textbf{Re-CNOT to RELOCATE Cost Ratio ($\alpha$).} The Re-CNOT to RELOCATE cost ratio depends on the hardware. 
Figure~\ref{fig:alpha} shows the teleportation count of a QAOA 3-regular program on $2\times2$ DQC relative to QuComm for various Re-CNOT to RELOCATE cost ratios. \design{} consistently reduces $T_{\text{eff}}$.

\begin{figure}[htpb]
\centering
\includegraphics[width=\columnwidth]{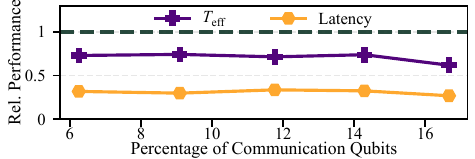}
\caption{Performance comparison using a 120-qubit QAOA 3-regular program on $2\times2$ DQC across various percentages of communication qubits. Values are normalized to QuComm.}
\label{fig:commbudget}
\end{figure}

\noindent \textbf{Percentage of Communication Qubits.} By default, IRIS reserves $10\%$ of qubits per chip for communication because non-local CNOT count is often much lower than local CNOTs. Figure~\ref{fig:commbudget} compares the performance of a 120-qubit QAOA 3-regular program on $2\times2$ DQC with increasing percentages of communication qubits. \design{} consistently outperforms.

\subsection{Impact of EPR Pair Generation Latency}
\begin{figure}[hbtp]

    \centering
\includegraphics[width=\linewidth]{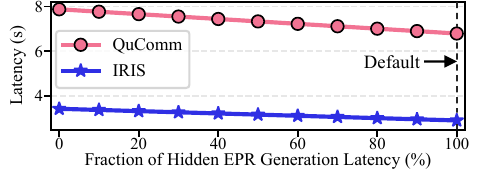}  
    \caption{Latency of a 240-qubit QAOA-FC program on $3$$\times$$3$ DQC with increasing fraction of hidden EPR generation latency. \design{} consistently outperforms QuComm.}
    \label{fig:epr_latency}
\end{figure}

We assume EPR pair generation can be fully overlapped with atom movements, but hardware limitations may sometimes prevent this.
For example, Rb atoms take 2.67$\times$ longer for EPR pair generation than Yb atoms~\cite{aqua2025mode,Li2024Highrate}.
Figure~\ref{fig:epr_latency} shows the latency of a 240-qubit QAOA-FC on a 3$\times$3 DQC as EPR pair generation latency hidden varies from 0\% (no hiding) to 100\% (fully hidden). \design{} consistently outperforms QuComm and is more advantageous when this latency cannot be hidden, as its schedule has fewer teleportations and higher concurrency. 

\subsection{Impact on Quantum Error Correction (QEC)}
Fault-tolerant quantum computers map program variables to logical qubits encoded in quantum error correction (QEC) codes and translate instructions to logical operations first. These operations are then translated into low-level hardware compatible instructions (similar to \design{}). \design{} can be integrated into existing compilers that focus on logical synthesis, but developing an end-to-end fault-tolerant DQC compiler is beyond the scope of our paper. To show that optimized low-level scheduling can still benefit fault-tolerant systems, we compare the scheduling of \design{} against QuComm for three promising QEC codes on a 2$\times$2 DQC. 

As shown in Table~\ref{tab:QEC}, \design{} reduces $T_{\mathrm{eff}}$
by 9--42\% and logical cycle time (in milliseconds) by 51--65\% compared to QuComm. Logical cycle time is the time taken to extract parity information by running a syndrome extraction circuit on data qubits (that store information) and parity qubits (that detect errors) over multiple rounds. 
It dictates the runtime of programs.
In surface code~\cite{SurfaceCode_Kitaev2003} and color code~\cite{ColorCode_Bombin2006}, each parity qubit interacts with only up to four data qubits, resulting in fewer non-local CNOTs than in bivariate bicycle code~\cite{BBCode_Bravyi2024}, where a parity qubit interacts with up to six data qubits and 
there is a greater likelihood that the data qubits are spread across multiple chips.
The costs may be further reduced with QEC-specific mappers~\cite{chipmunq}. We leave it for future work.

\begin{table}[htp]
\centering
\caption{Performance of a logical cycle on a 2$\times$2 DQC. }
\renewcommand{\arraystretch}{1.0}
\setlength{\tabcolsep}{3pt}
\label{tab:QEC}
    \begin{center}
    \begin{small}
    \begin{tabular}{|c||c|c|c|c|}
    \hline
    \multirow{2}{*}{Code} 
    & \multicolumn{2}{c|}{QuComm} & \multicolumn{2}{c|}{\design{}} \\
    \cline{2-5}
    \cline{2-5}
        & $T_\text{eff}$ & $L_{\text{cycle}}$ & $T_\text{eff}$ & $L_{\text{cycle}}$   \\
    \hline    
    \hline
          Bivariate Bicycle  [[72,12,6]] & 2566 & 1192 & 2337 & 588 \\\hline
          Color [[61,1,9]]              & 1106 & 1878 &  797 & 722 \\\hline
          Surface [[49,1,7]]            &  589 &  738 &  343 & 257 \\\hline\hline
          \textbf{Relative} & 1.00 & 1.00 & \textbf{0.73} & \textbf{0.40}\\\hline
    \end{tabular}
    \end{small}
    \end{center}
\end{table}

\section{Complexity Analysis}
We analyze \design{}'s complexity assuming a maximum block size $B$.
\textit{First}, BFA searches up to $B$ gates and fuses blocks by again searching up to $B$ gates per iteration. Over an entire program, with at most $\frac{V}{B}$ blocks, BFA has a complexity of $C_{\text{BFA}}\!\!=\mathcal{O}(V\!\cdot\!B)$. \textit{Second}, UMS schedules one block at a time while preserving up to $w$ solutions. To schedule each non-local gate, UMS considers up to $N$ chips.  If EPR release is needed, UMS additionally considers which qubit to move and to which chip. Thus, the complexity of scheduling a non-local gate is $C_{\text{gate}}=\mathcal{O}(N\log N + N\cdot n_{\text{qubit}})$. 
For the block's $i^{\textit{th}}$ non-local gate, \ums{} evaluates at most $|G|\cdot (B-i)$ future gates in the scheduling group.
As a block contains at most $B$ non-local gates and UMS preserves up to $w$ solutions, the complexity of UMS per block is $C_{\text{\ums{}}}=\mathcal{O}(w\cdot |G|\cdot B^2\cdot C_{\text{gate}})$. \design{} iterates \ums{} until all $V$ program gates are scheduled. The number of iterations is at most $\frac{V}{B}$.  
\textit{Lastly}, \ees{} identifies the set of operations whose data dependencies are resolved, which takes $|G| \cdot B$. For these operations, \ees{} linearly searches the previous operations, and its complexity is $C_{\text{\ees{}}} \!\!=\!\!\mathcal{O}(|G|^2 \cdot B^2)$.
The overall complexity of \design{} is given in Equation~\eqref{eq:complex}.
\begin{equation}
\label{eq:complex}
\resizebox{0.90\hsize}{!}{
$C_{\mathrm{\design{}}} =  \mathcal{O}\!( VB[ w\cdot |G|\cdot N\cdot \!(\log N + n_{\mathrm{qubit}}) + |G|^2 ] )$}
\end{equation}

When a block grows up to the entire program (worst case), the \design{} complexity scales quadratic in the program size. For a fixed maximum block size (typical case), solutions and group size, the complexity grows linearly with program size.

\vspace{0.05in}
\noindent \textbf{Scalability and Performance Comparison:}
We compare the  complexity and performance of \design{} and QuComm for QAOA 3-regular programs with three configurations. Table~\ref{tab:qaoatimecomplexity} shows the memory requirement, compiler runtime, effective teleportation count, and program latency for \design{} relative to QuComm. 
Although \design{} takes longer than QuComm to compile, its schedule has much lower execution time which leads to a reduced end-to-end program runtime because quantum programs run for thousands of shots. For example, compiling a 1100-qubit program with 3.3K CNOTs using QuComm takes 2 minutes and produces a schedule of 785ms latency. Assuming 50K shots, the total runtime of the program is 2\,minutes + 50K $\times$ 785\,ms = 656\,minutes. In contrast, \design{} takes 150\,minutes to compile but produces a schedule with 307\,ms latency, leading to a total runtime of 150\,minutes + 50K $\times$ 307\,ms = 406\,minutes, 38\% lower than QuComm.

\begin{table}[htb]
\begin{center}
\caption{Complexity analysis and performance of \design{} relative to QuComm for a program executing 50K shots. } 
\setlength{\tabcolsep}{0.8mm} 
\renewcommand{\arraystretch}{1.01}
\label{tab:qaoatimecomplexity}
\begin{small}
\begin{tabular}{ |c|c||c|c||c|c|c|} 
\hline
DQC & Program & \multirow{2}{*}{Memory}  & \multirow{2}{*}{$T_\text{eff}$} & Compile& Schedule& Program \\
Arch & Qubits &  & & Time & Latency & Runtime \\ \hline\hline
2$\times$2 & 500 & 2.08$\times$ & 0.83$\times$ &  363$\times$ & 0.38$\times$ & 0.80$\times$\\
\hline
2$\times$3 & 800 & 2.89$\times$ & 0.83$\times$  & 201$\times$ & 0.35$\times$ & 0.61$\times$ \\
\hline
3$\times$3 & 1100 & 2.53$\times$ & 0.85$\times$ & 74$\times$ & 0.39$\times$ & 0.62$\times$ \\
\hline
\end{tabular}
\end{small}
\end{center}
\label{tab:complexity}
\end{table}
\section{Related Work}
\label{sec:related_work}
 
We compare and contrast \design{} with prior DQC compilers.

\vspace{0.05in} 
\noindent
\textbf{DQC Mappers.} DQC mappers assign program qubits to compute qubits to minimize non-local gates.
{Static partitioning}~\cite{daei2020optimized,Cambiucci2023Hypergraphic,sundaram2021efficient,PARK1995899} computes a single assignment at compile time.
{Dynamic partitioning}~\cite{nikahd2021automated,baker2020cf,burt2024generalised,kaur2025optimized,chipmunq} instead divides a program into segments and reconfigures qubit layout per segment, incurring extra RELOCATEs at segment boundaries for fewer non-local gates within each segment.
These mappers provide better qubit layout but leave non-local gates for the scheduler.
\design{} focuses on instruction scheduling and is orthogonal to mapping. \design{} is compatible with both types of mapping policies and outperforms consistently. 

\vspace{0.05in}
\noindent\textbf{DQC Schedulers.}
Ferrari et al.~\cite{ferrari2020} independently schedule each non-local gate. AutoComm groups non-local gates that share a qubit and span two chips into a block~\cite{DQC-Wu2022AutoComm}.    QuComm~\cite{DQC-Wu2023QuComm} extends blocks across multiple chips under EPR capacity.
These methods have limited lookahead capabilities and \design{} outperforms them.
Cuomo et al.~\cite{cuomo2023optimized} formulate non-local gate scheduling as an ILP, but support only Re-CNOTs. 
As Re-CNOT performs a CNOT between two chips without relocating any qubit,  each qubit stays in its initial chip for the program. Thus, this compiler misses opportunities to reduce teleportations through qubit displacements.
\design{} overcomes this drawback. Promponas et al.~\cite{promponas2025compiler} use deep reinforcement learning for scheduling, but their method does not scale beyond 50 gates. In contrast, \design{} scales to large programs.
AdaptDQC adopts circuit cutting for non-local gate scheduling~\cite{Xiang2025AdaptDQC}.
However, the post-processing cost of circuit cutting grows exponentially in the number of non-local gates, limiting scalability. 
SwitchQNet~\cite{zhang2025switchqnet} reduces dynamic switching overheads in the all-to-all connected DQCs~\cite{cisco_arch1,cisco_arch2,cisco_arch3,cisco_arch4} rather than teleportations, which is orthogonal to \design{}.

\clearpage
\section{Conclusion}

We propose \textit{\design{}}, a compiler that reduces teleportation costs in Distributed Quantum Computers (DQCs).
\textit{\design{}} employs two features. \textit{First}, \textit{\underbar{U}tility-Driven Lookahead with \underbar{M}ulti- Candidate Block} \textit{\underbar{S}cheduling (\ums{})} to schedule operations by (1)~using a utility-driven lookahead
window that only considers useful future blocks and (2)~retaining multiple candidate schedules to prevent early commitment to sub-optimal schedules. \textit{Second}, IRIS employs \textit{\underbar{E}PR-Capacity-Aware \underbar{E}arly \underbar{S}cheduling (\ees{})} to exploit available EPR resources to initiate the execution of future gates and their teleportations early. Our evaluations show that \design{} reduces teleportations by $34\%$ on average and by up to $65\%$, while reducing latency by $2.0\times$ on average and by up to $2.9\times$ compared to the baseline.

\section*{Acknowledgments}
We thank Hezi Zhang for her help in implementing QuComm on our end. We thank Avinash Kumar, Sourish Wawdhane, Ravi Ghadia, Dongwhee Kim, and Reza Nejabati for generic discussions. We also thank the reviewers of EuroSys 2027 for their feedback and our shepherd for helping us improve the paper. Last but not the least, we thank the generous support from Cisco and Cockrell School of Engineering at UT Austin.

\appendix
\section{EPR Pair Generation Process}
\label{appendix:epr_generation}
\begin{figure}[htb] 
\centering
\includegraphics[width=\columnwidth]{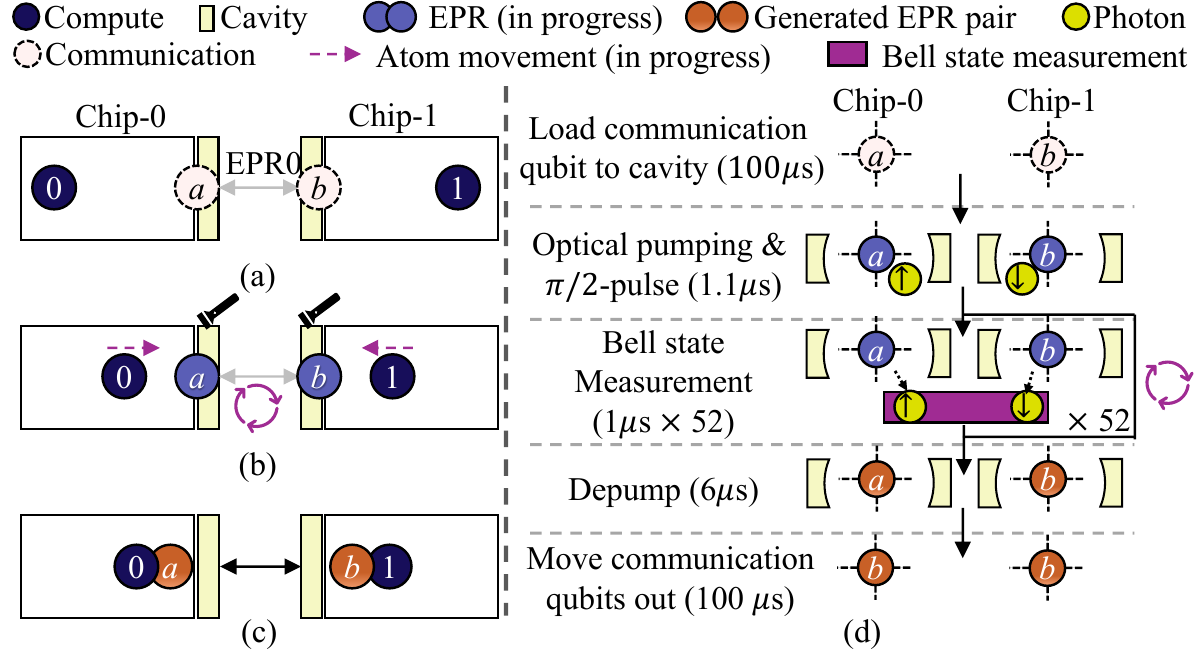} 
\caption{Overview of EPR pair generation.}
\label{fig:on_demand_epr}
\end{figure}

We explain EPR generation using Figure~\ref{fig:on_demand_epr}, based on the protocol by Li et al.~\cite{Li2024Highrate}.
A non-local CNOT requires an EPR pair, qubit movements, and local operations. 
To perform a CNOT between qubits $0$ and $1$, communication qubits $a$ and $b$ form an EPR pair. Qubits $0$ and $1$ move to the interaction range of $a$ and $b$ respectively, followed by CNOT $0$, $a$ and CNOT $b$, $1$ to form an EPR pair. The communication qubits are loaded into cavities and optically pumped. The photons emitted from the spin-up ($|\!\!\uparrow\rangle$) or spin-down state ($|\!\!\downarrow\rangle$) are directed to Bell state measurement (BSM) to check for a successful entanglement. This process is probabilistic with a success rate, $P$, of 12.5\%. To produce an EPR pair with a fidelity of 99.9\%, we use a repeat until success approach with $N$ attempts. The probability of successful entanglement after $N$ attempts is given by (1-(1-$P)^N$). For it to be 99.9\%, we need at least 52 attempts. Next, the qubits are depumped and moved out of the cavities. Based on device studies~\cite{bluvstein2022quantum,Li2024Highrate}, EPR pair generation takes 259$\upmu$s while moving atoms takes 300$\upmu$s. This allows us to fully hide the EPR generation latency.

\section{Protocol for Non-Local CNOTs}
\label{appendix:teleportations}
\begin{figure}[htpb]
\centering
\includegraphics[width=.98\columnwidth]{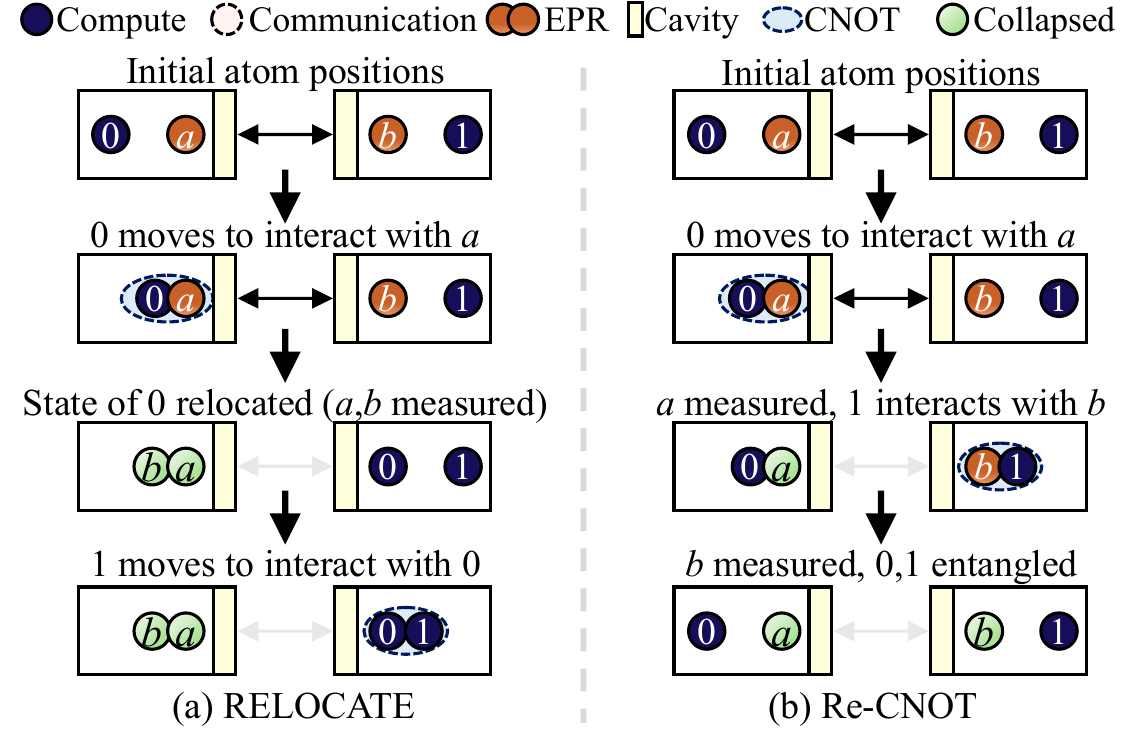}
\caption{Implementation of the non-local CNOT.}\label{fig:nonlocal_cnots}
\end{figure}

We explain the detailed execution of non-local CNOTs in neutral-atom DQCs.
Figure~\ref{fig:nonlocal_cnots}(a) shows a RELOCATE. Qubit $0$ moves within the interaction range of qubit $a$ and CNOT $0$, $a$ is executed. State teleportation is used during which qubits $a$ and $b$ collapse after measurement. CNOT $0$, $1$ is performed by bringing qubit $1$ within interaction range of qubit $0$.
Figure~\ref{fig:nonlocal_cnots}(b) shows a Re-CNOT operation. Qubit $0$ is moved within the interaction range of qubit $a$ and CNOT $0$, $a$ is executed. Gate teleportation is used here which measures qubit $a$ and prepares qubit $b$ as a proxy for the remote CNOT. Next, qubit $1$ is moved within the interaction range of $b$,  CNOT $b$,$1$ is executed, and $b$ collapses after measurement.

\section{\design{} Algorithm}\label{appendix:algorithm}
Algorithm~\ref{algo:iris} gives an overview of the \design{} algorithm.
\begin{small}
\begin{algorithm}[b]
\caption{\design{}}\label{algo:iris}
\begin{algorithmic}[1]
\Input{Program,  Qubit Mapping ($M$),  EPR Capacity ($D$) }
\Output{Schedule ($\textrm{Sched}$)}
\State \textbf{function} \design{}(Program, $M$, $D$)
\State \quad $\mathrm{DAG} \leftarrow \textit{Build\_DAG}(\mathrm{Program})$
\State \quad $\mathcal{B} \leftarrow \textit{Block\_Formation}(\mathrm{DAG}, M, D)$
\State \quad $\mathcal{T} \leftarrow \textit{Initialize\_Solution\_Tree}(M, D)$
\State \quad \textbf{while} not all blocks in $\mathcal{B}$ are scheduled \textbf{do}
\State \quad \quad $C \leftarrow$ next unscheduled block in $\mathcal{B}$
\State \quad \quad $G \leftarrow \textit{Utility\_Driven\_Lookahead\_Window}(C, \mathcal{B})$
\State \quad \quad $\mathcal{T} \leftarrow \textit{UMS}(C, G, \mathcal{T}, M, D)$
\State \quad $\mathrm{Sched} \leftarrow \textit{Best\_Schedule}(\mathcal{T})$
\State \quad Sched $ \leftarrow \textit{\ees{}}$(Sched)
\State \quad \textbf{return} Sched
\end{algorithmic}
\end{algorithm}
\end{small}

\section{Artifact Appendix}

\subsection{Abstract}
This artifact comprises a codebase and dataset. The codebase contains the DQC mappers and DQC schedulers. For the dataset, benchmark circuits, DQC mapping, the generated DQC schedules, and the raw results reported in this paper are included. 
The authors provide two reproduction workflows: a \emph{data-only} workflow is for the evaluators who use a small machine. It provides the paper's tables and figures directly from the \design{} dataset within an hour. A \emph{full} workflow that re-runs all compilations and reproduces the paper results in about 7 days on a 160-core server.

\subsection{Artifact check-list (meta-information)}
{\small
\begin{itemize}[leftmargin=2pt, itemindent=2pt, labelsep=3pt,itemsep=0pt]
 \item {\bf Algorithm:} QuComm (baseline) and \design{} (ours).
 \item {\bf Program:} Eight benchmarks (BV, QAOA-3reg, QAOA-FC, QFT, QuGAN, QV, Shor, and VQE) and three QEC codes (Bivariate Bicycle, Color, and Surface); all circuits are included.
 \item {\bf Data set:} \design{}-dataset (1.4\,GB archive on Zenodo) contains benchmark circuits, mapper outputs, DQC schedules, and the raw results.
 \item {\bf Run-time environment:} Linux; Python~3.9 Conda environment  (\texttt{environment.yml} provided).
 \item {\bf Hardware:} Any Linux machines ($\geq$16\,GB RAM) for the data-only workflow; a multi-core server is recommended for the full re-run (we use 160-cores machine).
 \item {\bf Metrics:} Effective teleportation count and program latency.
 \item {\bf Output:} CSV/LaTeX tables and PDF figures matching the paper.
 \item {\bf How much disk space required (approximately)?} About 40\,GB
 (1.4\,GB dataset; the rest for the full re-run).
 \item {\bf How much time is needed to prepare workflow (approximately)?}  Less than one hour.
 \item {\bf How much time is needed to complete experiments (approximately)?}  Data-only workflow takes under one hour, while full re-run takes about 7 days on a
 160-core machine.
 \item {\bf Publicly available?} Yes, codebase is available.
 \item {\bf Code licenses (if publicly available)?} MIT.
 \item {\bf Data licenses (if publicly available)?} MIT.
 \item {\bf Archived (provide DOI)?}
 10.5281/zenodo.22152933 ({\color{blue}{\href{https://zenodo.org/records/22152933}{Zenodo}}}).
\end{itemize}
}

\subsection{Description}
\noindent \textit{1) How to access:} The codebase, mapping generated by the mappers, schedule generated by the schedulers, benchmark programs and the results of raw data are available on {\color{blue}{\href{https://zenodo.org/records/22152933}{Zenodo}}}.

\vspace{0.05in}
\noindent \textit{2) Hardware dependencies:}
No special hardware and no GPU is required. The data-only workflow runs on any Linux/macOS machines with $\geq$16\,GB RAM. For the full re-run, the authors recommend using a compute cluster or a set of servers, because the time required for reproduction grows linearly in the number of cores. The authors use a 160-core machine.

\vspace{0.05in}
\noindent \textit{3) Software dependencies:} Conda (Miniforge/Anaconda). A \texttt{setup.sh} file creates the \texttt{iris-ae} Python environment from the environment file \texttt{environment.yml}.

\vspace{0.05in}
\noindent \textit{4) Data sets:} is archived on Zenodo. \texttt{IRIS-dataset.tar} file contains the benchmark circuits, DQC mappings, the generated schedule, and the raw results.

\subsection{Installation}
Run the following commands to set up:
\begin{Verbatim}[
  fontsize=\small,
  breaklines=true,
  breakanywhere=true,
  backgroundcolor=codegray,
  backgroundcolorpadding=3pt
]
bash setup.sh && conda activate iris-ae
\end{Verbatim}

\subsection{Experiment workflow}
The artifact supports two workflows that share the same output format.

\vspace{0.05in}
\noindent \textit{1) Data-only (Less than an hour):} This workflow regenerates every table and figure directly from the released schedules and results,
without re-running the compiler:

\begin{Verbatim}[
  fontsize=\small,
  breaklines=true,
  breakanywhere=true,
  backgroundcolor=codegray,
  backgroundcolorpadding=3pt
]
bash scripts/get_data_all.sh --from_dataset
\end{Verbatim}

\noindent \textit{2) Full re-run ($\sim$7 days on 160 cores):} This workflow rerun every scheduling experiment from the released mappings, then regenerate the tables and figures from the new results:

\begin{Verbatim}[
  fontsize=\small,
  breaklines=true,
  breakanywhere=true,
  backgroundcolor=codegray,
  backgroundcolorpadding=3pt
]
bash scripts/reproduce_all.sh
bash scripts/get_data_all.sh --from_results
\end{Verbatim}

\subsection{Evaluation and expected results}
The workflows generate experimental results. The tables are written as CSV/Latex files under \texttt{data\_generator/output/} and the figures as PDFs under \texttt{figures/}. The workflows reproduce the following tables and figures:
{
\begin{itemize}[leftmargin=2pt, itemindent=2pt, labelsep=3pt,itemsep=0pt]
 \item {\bf Tables:} \ref{tab:tab-1}, \ref{tab:tab-2},     \ref{tab:main_table_2x2}, \ref{tab:main_table_others}, \ref{tab:QEC},    and \ref{tab:qaoatimecomplexity}\ignore{, and \ref{tab:latency_compare}}.
 \item {\bf Figures:} \ref{fig:ums_blockid_teff}, \ref{fig:ums_blockid_future_teff},       \ref{fig:results_ees}, \ref{fig:each_optimization}, \ref{fig:ums_variants},  
 \ref{fig:result_scaling},    \ref{fig:results_beamwidth}, \ref{fig:alpha}, \ref{fig:commbudget},   and  \ref{fig:epr_latency}\ignore{and \ref{fig:fidelity_breakdown}}.
\end{itemize}
}
Every reported trend and relative improvement are expected to match the reported results; compile times may differ across execution environments (library versions and machines).

\subsection{Experiment customization}
Experiments can be customized with command-line arguments. 
For example, a single (benchmark, architecture, mapper) combination can be run with the following line:
\begin{Verbatim}[
  fontsize=\small,
  breaklines=true,
  breakanywhere=true,
  backgroundcolor=codegray,
  backgroundcolorpadding=3pt
]
bash scripts/run_all_variants.sh bv_n120 F120 ILP
\end{Verbatim}
In addition, custom architectures can be specified as \texttt{S<qubit s/chip>C<EPR/link>-<rows>x<cols>}, and a new benchmark can be added as OpenQASM~2 files under \texttt{bench/}.

\subsection{Methodology}
Submission, reviewing and badging methodology:
{\small
\begin{itemize}[leftmargin=2pt, itemindent=2pt, labelsep=3pt,itemsep=0pt]
  \item \url{https://www.acm.org/publications/policies/artifact-review-and-badging-current}
  \item \url{https://cTuning.org/ae}
\end{itemize}
}

\newpage
\bibliographystyle{unsrt}
\bibliography{reference}
\end{document}